\documentclass[sigconf, nonacm]{acmart}
\usepackage{xcolor}
\usepackage{amsmath,amsfonts}
\usepackage{algorithmic}
\usepackage{graphicx}
\usepackage{capt-of}
\usepackage{booktabs}
\usepackage{varwidth}
\newsavebox\tmpbox
\usepackage{enumitem}
\usepackage{multirow}
\usepackage{subcaption}
\usepackage{xspace}
\usepackage{placeins}
\usepackage{lscape}
\usepackage{pifont}
\usepackage{url}
\usepackage{color}
\definecolor{lightgray}{gray}{0.9}

\newcommand\greybox[1]{%
  \par\noindent\colorbox{lightgray}{%
    \begin{minipage}{0.47\textwidth}#1\end{minipage}%
  }%
  \vskip\baselineskip%
}

\AtBeginDocument{%
  }

\setcopyright{acmlicensed}
\copyrightyear{2018}
\acmYear{2018}
\acmDOI{XXXXXXX.XXXXXXX}

\acmConference[Conference acronym 'XX]{Make sure to enter the correct
  conference title from your rights confirmation emai}{June 03--05,
  2018}{Woodstock, NY}

\acmISBN{978-1-4503-XXXX-X/18/06}

\begin{document}

\title{Characterizing User Behavior: The Interplay Between Mobility Patterns and Mobile Traffic}


\author{Anne Josiane Kouam}
\affiliation{%
  \institution{TU Berlin, Germany}
  \city{}
  \country{}}
\email{}

\author{Aline Carneiro Viana}
\affiliation{%
  \institution{Inria, France}
  \city{}
  \country{}
}

\author{Mariano G. Beiró}
\affiliation{%
 \institution{CONICET, Argentina}
 \city{}
 \state{}
 \country{}}

\author{Leo Ferres}
\affiliation{%
  \institution{	Universidad del Desarrollo, Chile}
  \city{}
  \state{}
  \country{}}

\author{Luca Pappalardo}
\affiliation{%
  \institution{CNR, Italy}
  \city{}
  \state{}
  \country{}}
\email{}



\renewcommand{\shortauthors}{Trovato et al.}

\begin{abstract}
  Mobile devices have become essential for capturing human activity, and eXtended Data Records (XDRs) offer rich opportunities for detailed user behavior modeling, which is useful for designing personalized digital services. Previous studies have primarily focused on aggregated mobile traffic and mobility analyses, often neglecting individual-level insights. This paper introduces a novel approach that explores the dependency between traffic and mobility behaviors at the user level. By analyzing 13 individual features that encompass traffic patterns and various mobility aspects, we enhance the understanding of how these behaviors interact. Our advanced user modeling framework integrates traffic and mobility behaviors over time, allowing for fine-grained dependencies while maintaining population heterogeneity through user-specific signatures. Furthermore, we develop a Markov model that infers traffic behavior from mobility and vice versa, prioritizing significant dependencies while addressing privacy concerns. Using a week-long XDR dataset from 1,337,719 users across several provinces in Chile, we validate our approach, demonstrating its robustness and applicability in accurately inferring user behavior and matching mobility and traffic profiles across diverse urban contexts.
\end{abstract}



\received{20 February 2007}
\received[revised]{12 March 2009}
\received[accepted]{5 June 2009}

\maketitle

\section{Introduction}
As mobile devices increasingly serve as proxies for human presence and activity, mobile datasets generated by these devices provide a unique opportunity to characterize and model user behavior in unprecedented detail. Among these datasets, eXtended Data Records (XDRs) stand out for their potential to provide the most comprehensive coverage of human interactions with mobile networks. XDRs are timestamped and georeferenced records of network mobile data sessions, generated by each mobile device interacting with the operator’s network. Unlike traditional datasets such as Call Detail Records (CDRs), which are limited to specific calls and text messages  event types, XDRs offer a much broader scope, capturing a wide array of network activities, from app usage to internet traffic. This extensive data richness not only enhances our understanding of user interactions with mobile networks but also provides a framework for analyzing underlying mobility patterns.

The ability of XDR data to capture both network usage and localization context has made it an acknowledged source of information for various studies, including human mobility analysis~\cite{Jiang:2013}, cellular traffic patterns~\cite{Eduardo:2015}, and network infrastructure usage~\cite{Ozturk:2021}. By analyzing these mobility patterns—such as the frequency and nature of location changes—we can gain valuable insights into user preferences, engagement levels, and behavioral tendencies. Such insights are vital for creating personalized services in a rapidly evolving digital landscape. Understanding when and where users are most active can inform adaptive content delivery, personalized recommendations, and targeted advertising strategies, ultimately enhancing user satisfaction and engagement. This intersection of mobility and mobile traffic is crucial for developing more tailored and responsive user experiences in the digital realm.

The interplay between mobile traffic and individual mobility patterns is crucial for understanding how users interact with digital platforms. Several studies have established significant correlations between mobile traffic and human mobility, primarily at an aggregate or city-wide scale. For instance, \cite{Wang:2019} characterized the geographical distribution of cellular traffic, revealing strong spatiotemporal dependencies even among distant cell towers due to transportation patterns across urban environments. Similarly, \cite{Wu_Jing:2018} demonstrated that cellular tower traffic can be indicative of land use, establishing profiles such as residential, transport, office, and entertainment zones based on traffic patterns in metropolitan areas. \cite{Chen:2015} further highlighted the spatial inhomogeneity of hourly traffic patterns between urban and rural regions, emphasizing how uneven traffic distribution reflects the underlying city infrastructure. 

While these studies provide valuable insights into the spatial and temporal dependencies of traffic at large scales, fewer have examined traffic and mobility behavior at the individual level. 
Some works, such as \cite{Alipour:2018} classified network devices into stationary ('cellos') and mobile ('flutes') categories, establishing that more mobile devices generate heavier traffic. \cite{Paul:2011} similarly found that subscribers with higher mobility produced more traffic in 3G networks, though their focus remained on network resource usage rather than deeper mobility dynamics. Despite these efforts, the granularity of individual behavior remains underexplored, as the uncovered correlations are not modeled, limiting the ability to reproduce or incorporate them into personalized insights, crucial for improving services like adaptive content delivery and traffic prediction.

\textit{This paper presents a novel approach to understanding the dependencies between traffic and mobility behaviors at the individual level.} Unlike existing studies that primarily focus on aggregated or city-scale analyses, we directly analyze these behaviors at the user level, revealing their potential for fine-grained behavioral modeling. By moving beyond correlations, which simply reflect statistical relationships, we focus on dependencies that capture how mobility directly influences traffic and vice versa. \textit{Our approach enhances the depth and accuracy of mobile network data interpretation by characterizing and modeling these dependencies, introducing a key inference mechanism that allows mobility behavior to be derived from traffic patterns and vice versa. This capability enables more accurate dataset integration and predictive modeling through a flexible, user-specific model that captures both spatial and traffic dynamics.}

We make the following contributions:
\begin{itemize}[leftmargin=*]
    \item \textbf{Comprehensive characterization (\S \ref{sec:contrib1}):} We provide a detailed analysis of the dependency between traffic and mobility behaviors through 13 individual features that encompass traffic patterns as well as spatial, structural, and social mobility aspects. This analysis incorporates more recent and nuanced profiling techniques that account for the exploration phenomenon in mobility, classifying users as \textit{regular, routiner,} or \textit{explorer}. As a result, this characterization enhances our understanding of how traffic and mobility patterns differ among various user types.
    
    \item \textbf{Advanced user modeling (\S \ref{sec:contrib2}):} We propose a framework modeling the dependencies between traffic and mobility, integrating both behaviors over time to capture the fine-grained dynamics inherent in XDR data. This model preserves population heterogeneity while accounting for individual differences through user-specific signatures, ensuring personalization and applicability across diverse urban environments.

    \item \textbf{Traffic-Mobility inference (\S \ref{sec:contrib3}):} Building on our proposed modeling, we investigate the reciprocal relationship between traffic and mobility behaviors. We develop a Markov model that leverages the identified dependencies, enabling traffic behavior to be inferred from mobility and vice versa. Unlike heavy deep-learning approaches that ingest raw data at the expense of generality and often raise privacy concerns due to data memorization, our model focuses on the most statistically significant dependencies. This ensures both privacy and adaptability across different contexts. The model not only generates traffic behaviors from mobility ones (and vice versa) but also assesses the likelihood of matching traffic and mobility behaviors, facilitating the creation of integrated datasets that fully capture both aspects.

    \item \textbf{Extensive validation:} Using a week-long XDR dataset comprising 1,337,719 users across several provinces in Chile, we validate our approach by demonstrating its ability to accurately infer user behavior. Additionally, our model can differentiate between likely and unlikely matches of mobility and traffic profiles, even when applied across different urban contexts, underscoring its robustness and general applicability. 
\end{itemize}
\vspace{-0.2cm}

Additionally, we review the background and related works in \textbf{\S \ref{sec:related_works}}. \textbf{\S \ref{sec:preliminaries}} outlines the methodology. Finally, we conclude in \textbf{\S \ref{sec:conclusion}}.

\section{Background and Related Works}
\label{sec:related_works}
Mobile traffic data and human mobility data each provide unique insights into human activity. 
While often studied separately, their strong interrelation has been demonstrated in various research. This section reviews the state-of-the-art studies on both data types and examines existing research on their correlation.

\vspace{0.13cm}
\noindent \textbf{Human Mobility data.} Human mobility is a well-researched area with significant value across several fields, including public health, sociology, transportation, and tourism~\cite{DeDomenico:2013, Koszowski:2019}. 
Studies have examined it from multiple angles, identifying key laws that govern movements~\cite{Gonzalez:2008, Rhee:2011} and creating models for simulating~\cite{Keramat:2016}, predicting~\cite{Wang_Jinzhong:2019}, and generating~\cite{Kobayashi:2023} mobility patterns at various scales. 
When analyzed at an individual level, human mobility provides more detailed insights by accounting for personal differences.
Research shows that individuals exhibit varied behaviors in their movement patterns~\cite{Pappalardo:2015, Scherrer:2018} 
with a recent study~\cite{Amichi:2020} identifying three mobility profiles
(i) \textit{Scouters}, more inclined to explore and discover new areas; (ii) \textit{Routiners}, who maintain a steady routine and rarely break their established patterns
; and (iii) \textit{Regulars}, with a moderate behavior balancing between explorations and revisits. 

\vspace{0.13cm}
\noindent \textbf{Mobile Traffic data} captures daily user activity on cellular networks valuable for tasks like network optimization and resource management. This study focuses on XDR traces, extensively studied in the literature for characterization, prediction~\cite{Aceto:2021, Trinh:2017, Wu_Jing:2018}, or generation~\cite{Kouam:2023} purposes.
Similar to human mobility data, individual-level analyses reveal distinct user profiles in data generation~\cite{Naboulsi:2014, Jing:2018, Eduardo:2015}. \cite{Eduardo:2015} distinguishes four profiles: \textit{Light Occasional (LO), Light Frequent (LF), Heavy Occasional (HO), and Heavy Frequent (HF)}. Light users generate up to 20GB per day, while Heavy users exceed this amount. "Occasional" and "Frequent" distinctions depend on the number of sessions users generate daily.

\vspace{0.13cm}
\noindent \textbf{Mobility and Traffic data dependency.}
Several studies have demonstrated the close relationship between mobile traffic and human mobility through statistical correlations \cite{Chen:2015, Wang:2019, Xu:2017, Alipour:2018, Merkebe:2013, Paul:2011}. The movement patterns of mobile network users—shaped by daily routines, points of interest, and behaviors—introduce significant temporal and spatial dynamics into mobile traffic data, which are crucial for understanding, modeling, and predicting cellular traffic at both large and fine scales. Additionally, the prevalence of network events among dispersed urban populations provides rich datasets that reflect underlying mobility patterns and urban dynamics, which are essential for effective city management and planning.

Existing studies on the correlation between mobility and mobile traffic have primarily focused on the city scale. For instance, \cite{Xu_Kai:2021} examined the geographic distribution of cellular traffic and uncovered strong spatiotemporal dependencies, while \cite{Chen:2015} identified spatial inhomogeneities in traffic patterns between urban and rural areas. Similarly, \cite{Merkebe:2013} illustrated how mobile traffic data can reveal city dynamics and infrastructure usage. However, while these studies offer valuable insights, they do not fully address the correlation between traffic and mobility at the individual level. Some works, such as \cite{Alipour:2018}, explored mobility-traffic correlations in WLAN settings at the device level, categorizing devices as "cellos" (stationary) and "flutes" (mobile), and found that cellos generated traffic more frequently, albeit in smaller volumes. Similarly, \cite{Paul:2011} investigated the relationship between subscriber mobility and traffic generation in 3G networks, concluding that more mobile users produced higher traffic volumes. 
While these approaches provide a preliminary understanding of how mobility correlates with traffic, a more detailed examination of the traffic-mobility interrelation at the individual behavior level remains largely unexplored, limiting the flexibility and potential of individual behavior profiling.

\vspace{0.13cm} \noindent \textbf{Positioning.} This paper introduces a novel approach that establishes a direct dependency between traffic and mobility behavior at the individual level. We explore the implications of this dependency for flexible user behavior modeling and the accurate inference of behavioral patterns in mobile networks.

\section{Preliminaries}
\label{sec:preliminaries}
This section presents an overview of our research approach. In \S \ref{subsec:overview}, we outline the methodology employed in this study, followed by a description of the dataset used for our analyses in \S \ref{subsec:data_desc}.

\subsection{Methodology overview}
\label{subsec:overview}

XDR data captures each mobile user's communication behavior across two dimensions along the \textit{time}: \textit{traffic}, and \textit{mobility}. A mobile user's daily activity, as recorded in an XDR dataset, can be represented as a temporal sequence of events $S^u = [e_1, e_2, \dots, e_n]$, where each event $e_n$ is a tuple $(t_n, v_n, l_n)$. In this tuple, $t_n$ represents the timestamp, discretized at a chosen granularity (e.g., 15 or 30 minutes), $v_n$ is the data volume generated during that time period, and $l_n$ indicates the corresponding location information.
\noindent We analyze the dependency between XDR traffic and mobility dimensions at an individual level according to the following methodology:
\begin{enumerate}[leftmargin=*]
\item In \textbf{\S \ref{sec:contrib1},} we identify and select a set of features for each dimension that characterize user behavior. We analyze this behavior based on the selected features 
including intersections and dependencies between features from different dimensions.
\item In \textbf{\S \ref{sec:contrib2}}, we introduce a novel user modeling capturing the dependency between the two dimensions allowing a finer understanding that is generalizable to multiple spatial contexts.
\item Building on this user modeling, we propose in \textbf{\S \ref{sec:contrib3}} a Markov model-based methodology allowing to infer a user’s mobility behavior from their traffic dimensions, and inversely. We chose a Markov model for its simplicity, ease of training, and explainability, making it lightweight and adaptable to a variety of contexts. Unlike deep-learning approaches that often compromise generalizability and raise privacy concerns through data memorization, the Markov model ensures both interpretability and flexibility across diverse applications.
\end{enumerate}

\subsection{Data Description}
\label{subsec:data_desc}
The dataset used in this study is an XDR mobile dataset, encompassing mobility and traffic information for a fully anonymized set of users over one regular week and covering the entire country of Chile at the time granularity, i.e., $t_{i+1}-t_i$, of 30 minutes.

\paragraph{Raw data description.} 
The mobility information \( l_n \) records the reference code of the most frequently visited antenna during the corresponding interval. This reference code must be matched with an additional dataset to obtain the corresponding geographical location.
The traffic information \( v_n \) corresponds to the total data volume generated by the user within each 30-minute interval. 
If a user does not generate any traffic during a time interval \(i\), \(v_i\) is set to 0, and no corresponding location \( l_i \) is recorded for that time slot. Although no explicit timestamps are provided, the records are ordered sequentially in 30-minute intervals, beginning from midnight on Sunday, allowing for temporal reconstruction. 

\paragraph{Data Preprocessing.} Chile is administratively divided into several hierarchical levels, with provinces as a key component, totaling 56 provinces. Provinces represent smaller administrative units within a region. For this study, we organize the data at the provincial level, creating multiple datasets—one for each province—allowing us to apply and validate the methodology across zones of varying scales and urban characteristics. A user is assigned to a province based on where they are most frequently observed during the week.

In this paper, due to space constraints, we present the results for four of the most populated provinces in Chile: Santiago, Elqui, Bio-Bio, and Copiapo. Table~\ref{tab:data-description} provides details on the number of users per province, the size, as well as the number of cell towers located within the geographical boundaries of each province. These provinces were chosen for their diversity in terms of urbanization profiles: Santiago, as the nation's capital, represents a highly urbanized and dense area, while Elqui and Bio-Bio offer insights into more moderate urban environments with distinct regional characteristics. Copiapo, though smaller, provides a contrasting profile with its unique rural-urban dynamics.

After matching the location reference codes in the raw dataset with the corresponding geographical coordinates, some location codes were found to be missing. To ensure data quality, users with more than 5\% missing location data were excluded. For users with less than 5\% missing locations, the missing data points were imputed based on the last known location. The final user counts after filtering are presented in the 2nd column of Table~\ref{tab:data-description}. Additionally, the 3rd column shows the number of users with a full sequence of $7\times48=336$ recorded events, having no inter-event time and no missing location data throughout the entire week.

\begin{table}[]
\centering
\caption{Dataset statistics for the provinces under study.}
\label{tab:data-description}
\resizebox{\columnwidth}{!}{%
\begin{tabular}{l|l|l|l|l|l|}
\cline{2-6}
 &
  \textbf{\begin{tabular}[c]{@{}l@{}}\#users b/f\\filtering\end{tabular}} &
  \textbf{\begin{tabular}[c]{@{}l@{}}\#users a/f\\filtering\end{tabular}} &
  \textbf{\begin{tabular}[c]{@{}l@{}}\#users w/\\full sequence\end{tabular}} &
  \textbf{\#cells}&
  \textbf{size} \\ \hline
Santiago       & 1,093,221 & 787326  & 21849  & 1536 & ~2,030$km^2$\\ \hline
Elqui          & 122,602   & 83041   & 2873   & 170 & ~16,895$km^2$ \\ \hline
Bio-Bio        & 65,049    & 38035   & 1106   & 87  & ~32,538$km^2$\\ \hline
Copiapo        & 56,847    & 3223    & 1115   & 71 &  ~14,987$km^2$\\ \hline
\textbf{Total} & 1,337,719 & 950,388 & 26,943 & 1864 & ~66,450$km^2$\\ \hline
\end{tabular}
}
\vspace{-0.7cm}
\end{table}

\section{Feature-based User Characterization}
\label{sec:contrib1}
We extract and compute several key features across mobility and traffic dimensions facilitating the identification of dependencies between them. The feature descriptions are provided in \S \ref{subsec:behavior_features}, followed by a detailed characterization in \S \ref{subsec:cor_analysis}.

\begin{table*}[]
\centering
\caption{Features description}
\label{tab:features_desc}
\resizebox{\linewidth}{!}{%
\begin{tabular}{|llll|l|l|}
\hline
\multicolumn{3}{|l|}{\textbf{Features}} &
  \textbf{Description} &
  \textbf{Equation (for $S^u$)} &
  \textbf{Step-level} \\ \hline
\multicolumn{1}{|l|}{\multirow{2}{*}{\textbf{Traffic}}} &
  \multicolumn{3}{l|}{\textit{avg. \#events per day} (Fig. \ref{fig:avgnev_user})} &
  $(\sum_{\{i \mid v_i \neq 0\}} 1)/N_{day}$ &
  / \\ \cline{2-6} 
\multicolumn{1}{|l|}{} &
  \multicolumn{3}{l|}{\textit{avg. session volume} (Figs \ref{fig:avg_volume}, \ref{fig:hour_volume_norm})} &
  $\frac{1}{n} \sum_{i=1}^n v_i$ & $trc_i \in \{l, m, h\}$ \\ \hline
\multicolumn{1}{|l|}{\multirow{12}{*}{\textbf{Mobility}}} &
  \multicolumn{1}{l|}{\multirow{4}{*}{\textbf{spatial}}} &
  \multicolumn{2}{l|}{\textit{avg. traveled  distance per slot} (Fig. \ref{fig:avg_distance})} &
  $\sum_{i=1}^{n-1} d_i$ with $d_i = |l_{i+1}-l_i|$ &
  $disc_i \in \{c, m, f\}$ \\ \cline{3-6} 
\multicolumn{1}{|l|}{} &
  \multicolumn{1}{l|}{} &
  \multicolumn{1}{l|}{\textit{Rg\_unique} (Fig. \ref{fig:rg_unique})} &
  \multirow{3}{*}{\begin{tabular}[c]{@{}l@{}}quantifies the spatial extent of users' movements:\\ higher values of \( r_g \) indicate that the user covers a large area, \\ while lower values suggest more localized movements.\end{tabular}} &
  $\sqrt{\frac{1}{n} \sum_{k=1}^n |l_k - l_{\text{unique}}|^2}, l_{\text{unique}} = \frac{1}{n} \sum_{k=1}^n l_k$ &
  / \\ \cline{3-3} \cline{5-6} 
\multicolumn{1}{|l|}{} &
  \multicolumn{1}{l|}{} &
  \multicolumn{1}{l|}{\multirow{2}{*}{\textit{Rg\_event} (Fig. \ref{fig:rg_event})}} &
   &
  \multirow{2}{*}{$\sqrt{\frac{\sum_{k=1}^n m_k \cdot (|l_k - l_{\text{event}}|)^2}{\sum_{i=1}^n m_i}}, \quad l_{\text{event}} = \frac{\sum_{k=1}^n m_k l_k}{\sum_{i=1}^n m_i}
    $} &
  \multirow{2}{*}{/} \\
\multicolumn{1}{|l|}{} &
  \multicolumn{1}{l|}{} &
  \multicolumn{1}{l|}{} &
   &
   &
   \\ \cline{2-6} 
\multicolumn{1}{|l|}{} &
  \multicolumn{1}{l|}{\multirow{6}{*}{\textbf{structural}}} &
  \multicolumn{1}{l|}{\textit{repetitiveness} (Fig. \ref{fig:rep_sta})} &
  Frequency of returns to previously visited locations &
  $(1 - n_{\text{unique}})/n$ &
  $rep_i=0/1$ \\ \cline{3-6} 
\multicolumn{1}{|l|}{} &
  \multicolumn{1}{l|}{} &
  \multicolumn{1}{l|}{\textit{stationarity} (Fig. \ref{fig:rep_sta})} &
  Ratio of user stays at the same location in consecutive intervals &
  $ \sum_{\{i \mid l_{i+1} = l_i\}} 1/(n-1)$ &
  $sta_i=0/1$ \\ \cline{3-6} 
\multicolumn{1}{|l|}{} &
  \multicolumn{1}{l|}{} &
  \multicolumn{1}{l|}{\textit{diversity} (Fig. \ref{fig:div_pre})} &
  Number of distinct sub-trajectories in a user trajectory &
  / &
  $\delta div_i$ \\ \cline{3-6} 
\multicolumn{1}{|l|}{} &
  \multicolumn{1}{l|}{} &
  \multicolumn{1}{l|}{\textit{predictability} (Fig. \ref{fig:div_pre})} &
  Entropy rate of the trajectory &
  / &
  / \\ \cline{3-6} 
\multicolumn{1}{|l|}{} &
  \multicolumn{1}{l|}{} &
  \multicolumn{1}{l|}{\textit{\#succ\_ret} (Fig. \ref{fig:mob_profiles})} &
  Number of successive returns &
  / &
  / \\ \cline{3-6} 
\multicolumn{1}{|l|}{} &
  \multicolumn{1}{l|}{} &
  \multicolumn{1}{l|}{\textit{\#succ\_expl} (Fig. \ref{fig:mob_profiles})} &
  Number of successive explorations &
  / &
  / \\ \cline{2-6} 
\multicolumn{1}{|l|}{} &
  \multicolumn{1}{l|}{\multirow{2}{*}{\textbf{social}}} &
  \multicolumn{1}{l|}{\textit{popularity influence} (Fig. \ref{fig:popularity})} &
  avg. visitation rate of a user's visited locations &
  $\frac{1}{n} \sum_{i=1}^{n} p(l_i, t_i)$ &
 $p(l_i, t_i)$ \\ \cline{3-6} 
\multicolumn{1}{|l|}{} &
  \multicolumn{1}{l|}{} &
  \multicolumn{1}{l|}{\textit{flow measurement} (Fig. \ref{fig:flow})} &
  avg. number of user with the same spatial movements &
   $\frac{1}{n} \sum_{i=1}^{n} flow(l_i, l_{i+1}, t_i)$ & 
   $flow(l_i, l_{i+1}, t_i)$
   \\ \hline
\end{tabular}%
}
\end{table*}

\subsection{Mobile user behavior features}
\label{subsec:behavior_features}

Table \ref{tab:features_desc} summarizes the features computed for each user, organized by XDR dimension (traffic and mobility) and categorized into spatial, structural, and social aspects for mobility features.

\vspace{0.13cm}
\noindent\textbf{Traffic Behavior.} We select well-known, simple features to characterize each user’s traffic behavior: the \textit{average number of events per day}, representing traffic frequency, and the \textit{average session volume}, capturing data usage per session. Based on these features, we derive a traffic profile for each user $P_u^{Tr} \in \{HO, HF, LO, LF\}$ (cf. \S \ref{sec:related_works}) through hierarchical clustering, as in \cite{Eduardo:2015}.

\vspace{0.13cm}
\noindent\textbf{Mobility Behavior.} We distinguish the following aspects:
\vspace{-0.13cm}
\begin{itemize}[leftmargin=*]
    \item \textbf{Spatial features} describe user mobility based on the geographical patterns of their movements. We consider the user's \textit{average traveled distance} and the \textit{radius of gyration}, a common metric in mobility analysis. The radius of gyration is computed in two forms: \textit{Rg\_unique}, considering each location once, and \textit{Rg\_event}, which weights locations based on visit frequency.
    
    \item \textbf{Structural features} capture the sequence patterns of user movements, focusing on the consistency and variability of their visits. Here, we use literature well-established metrics such as \textit{repetitiveness, stationarity, diversity}, and \textit{predictability}~\cite{mucceli:2016}. We calculate trajectory predictability using the Kontoyiannis entropy algorithm~\cite{Kontoyiannis:1998}. Additionally, we track \( \#succ_{ret} \) (successive returns) and \( \#succ_{expl} \) (successive explorations) to classify users into mobility profiles $P_u^{Mob} \in \{\textit{routiner, regular, scouter}\}$ using the clustering method from \cite{Amichi:2020}.
    
    \item \textbf{Social features} represent the influence of social interactions on mobility. We propose two metrics: (i) \textit{user popularity influence}, calculated as the average popularity of the locations visited by the user, where popularity \( p(l_i, t_j) \) is the number of unique users visiting location \( l_i \) at time \( t_j \). Additionally, (ii) we measure $flow(l_i, l_{i+1}, t_j)$ as the number of users following the same path from \( l_i \) to \( l_{i+1} \) during time slot $t_j$. Each user’s \textit{flow measurement} is the average of these flows across their entire trajectory.
\end{itemize}

\subsection{Feature-based analysis}
\label{subsec:cor_analysis}

This section leverages the features in \S \ref{subsec:behavior_features} in two key analyses: \textbf{(i)} Characterizing the global distribution of the population across traffic and mobility dimensions, and \textbf{(ii)} Examining how a user’s traffic profile is influenced by their mobility patterns, and inversely.

\begin{figure}[htbp]
    \centering
    \begin{subfigure}{0.44\columnwidth}
        \centering
        \includegraphics[width=\linewidth]{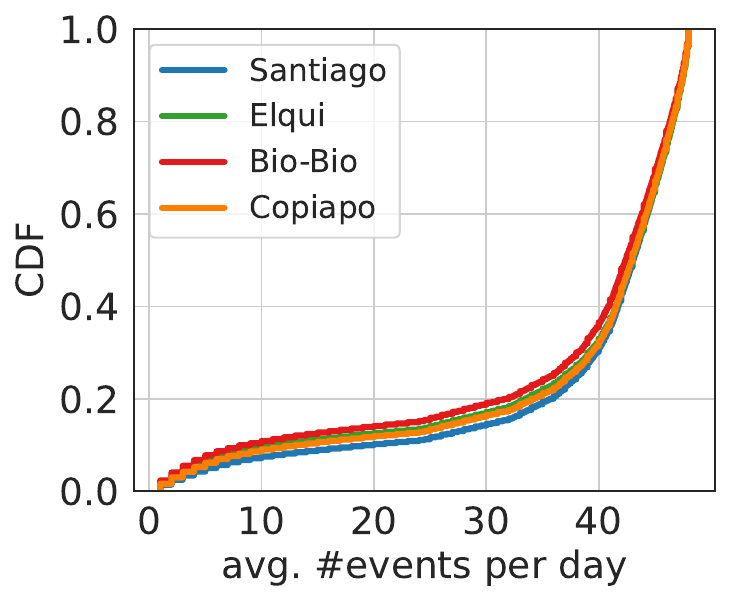}
        \caption{Number of events}
        \label{fig:avgnev_user}
    \end{subfigure}
    \hfill
    \begin{subfigure}{0.55\columnwidth}
        \centering
        \includegraphics[width=\linewidth]{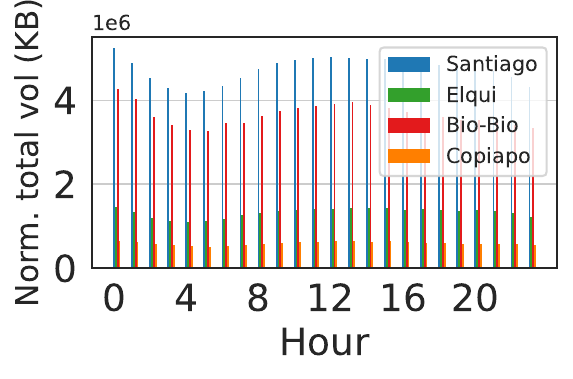}
        \caption{Hourly normalized volume}
        \label{fig:hour_volume_norm}
    \end{subfigure}
    \caption{Mobile users' traffic behavior distributions}
    \label{fig:traffic-behavior}
\end{figure}

\begin{figure}
    \centering
    \begin{subfigure}{0.32\linewidth}
        \centering
        \includegraphics[width=\columnwidth]{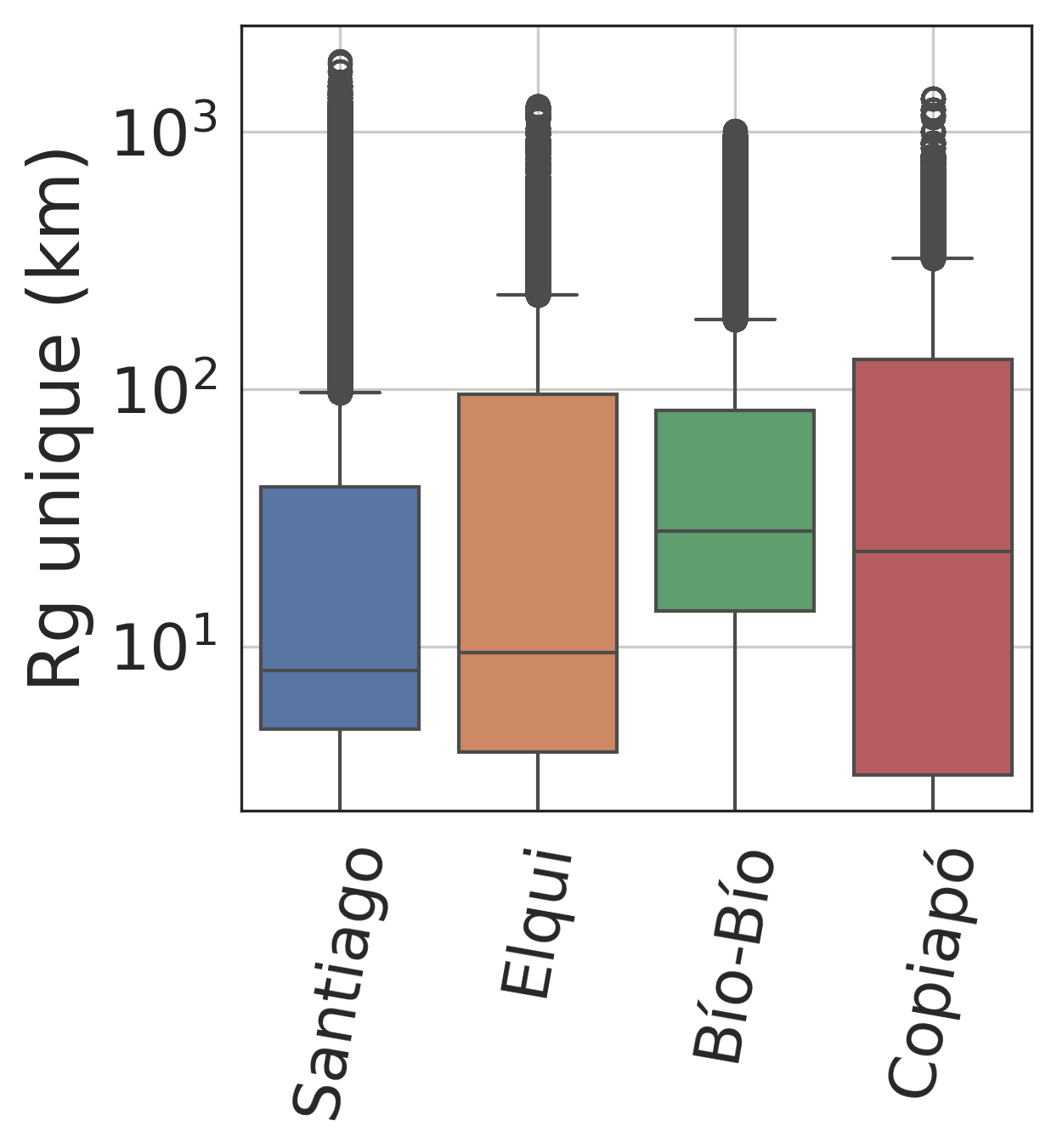}
        \caption{Radius of gyration unique}
        \label{fig:rg_unique}
    \end{subfigure}
    \hfill
    \begin{subfigure}{0.32\linewidth}
        \centering
        \includegraphics[width=\linewidth]{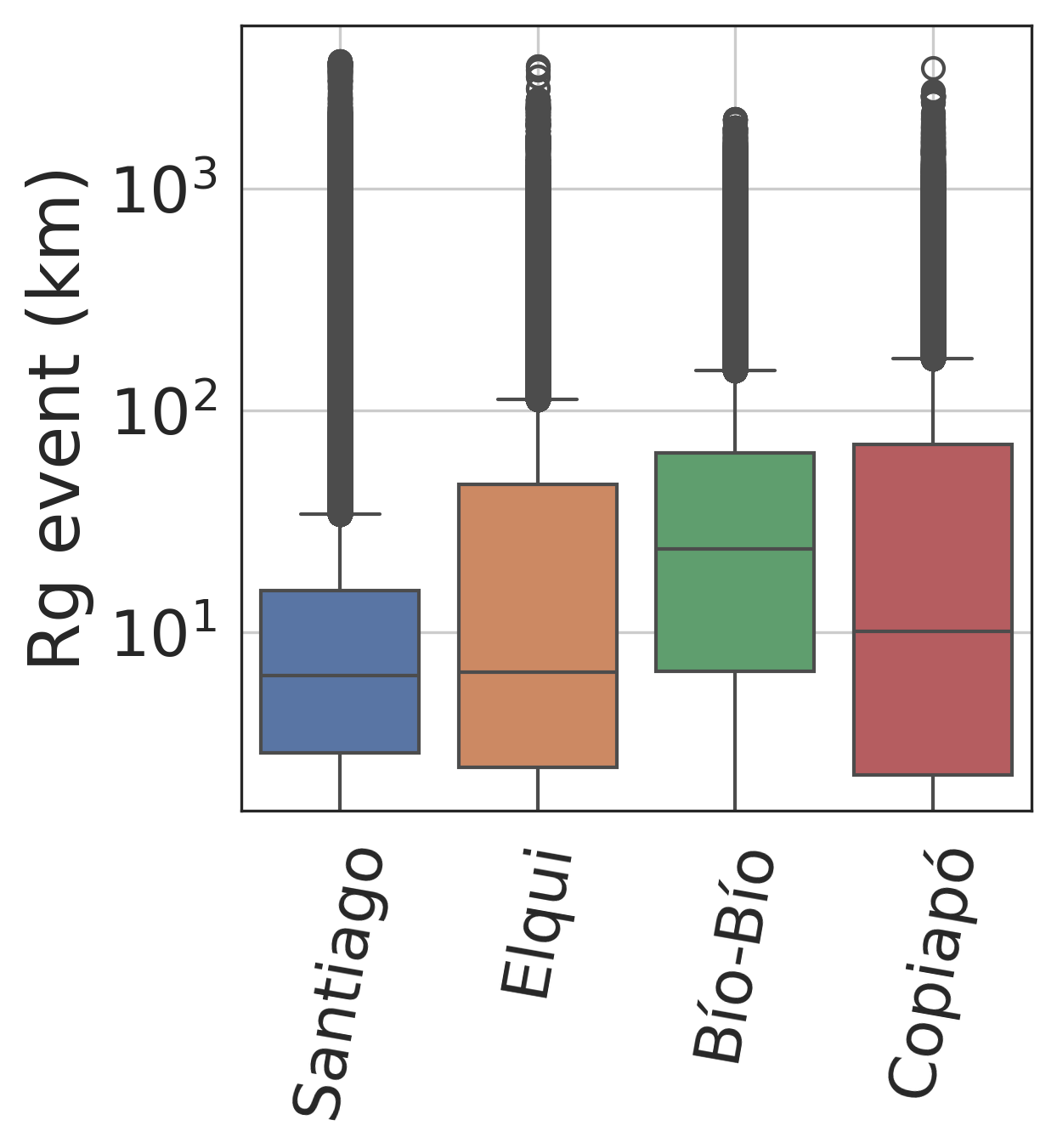}
        \caption{Radius of gyration event}
        \label{fig:rg_event}
    \end{subfigure}
    \hfill
    \begin{subfigure}{0.32\linewidth}
        \centering
        \includegraphics[width=\linewidth]{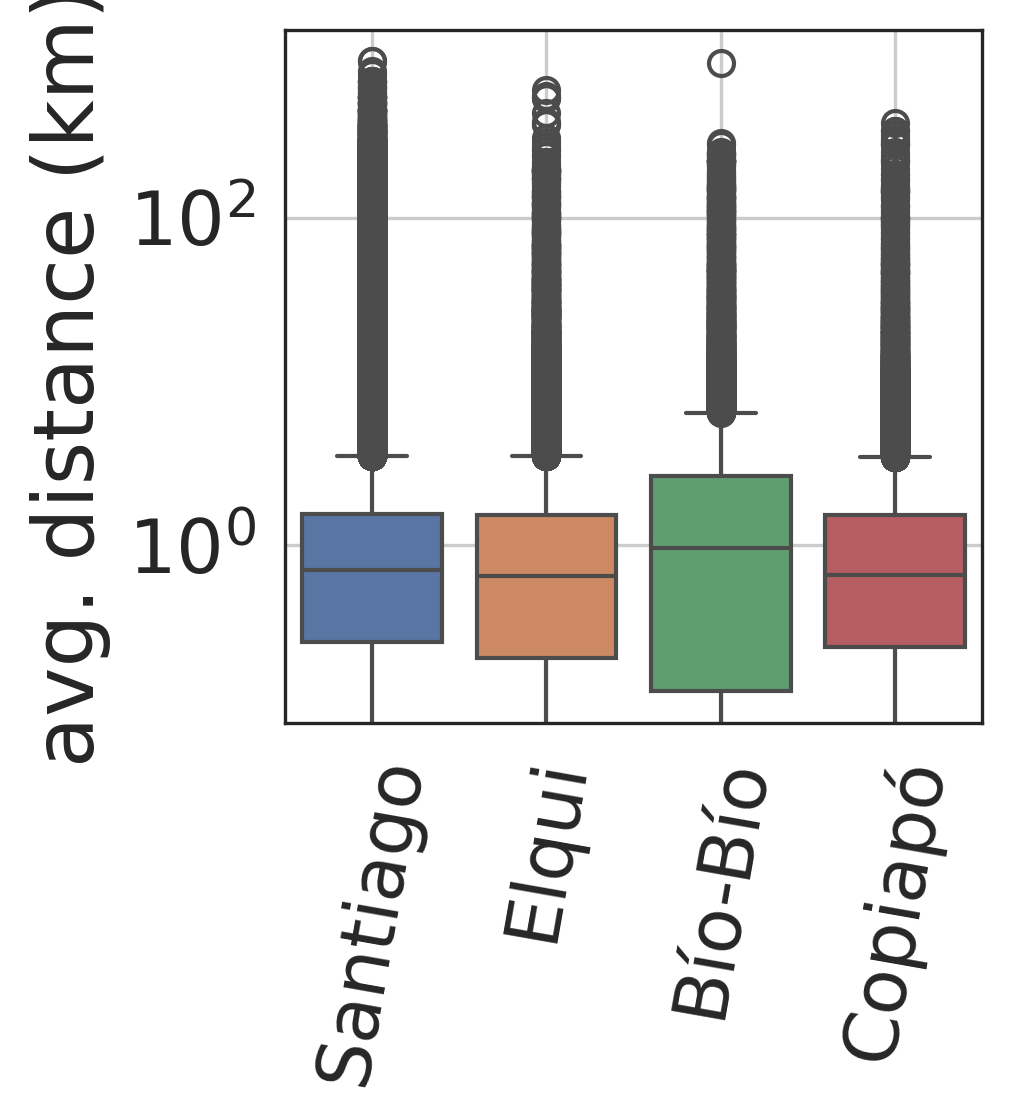}
        \caption{Avg. distance}
        \label{fig:avg_distance}
    \end{subfigure}
     \hfill
    \begin{subfigure}{0.48\linewidth}
        \centering
        \includegraphics[width=\linewidth]{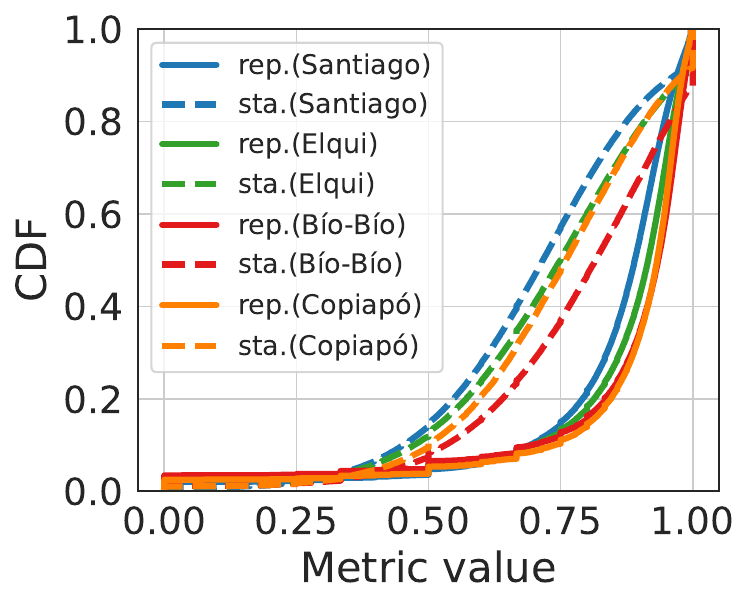}
        \caption{Repetitiveness (rep) and Stationarity (sta)}
        \label{fig:rep_sta}
    \end{subfigure}
     \hfill
    \begin{subfigure}{0.48\linewidth}
        \centering
        \includegraphics[width=\linewidth]{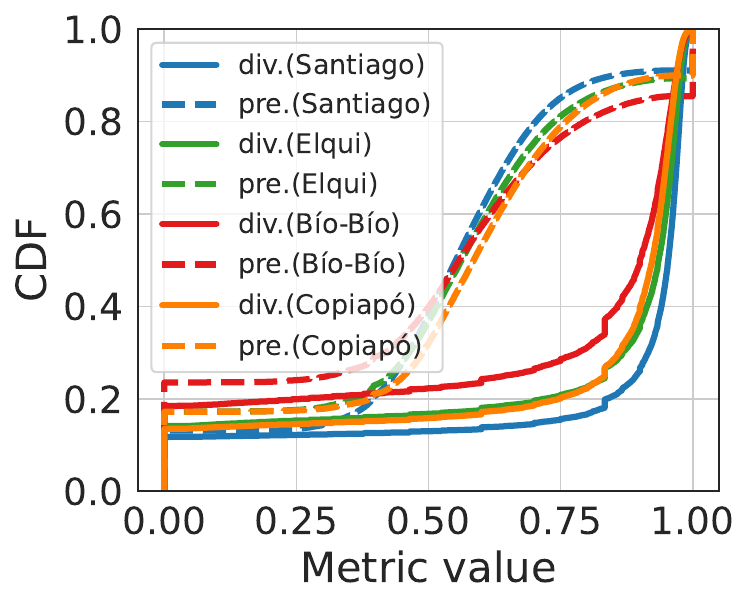}
        \caption{Diversity (div) and Predictability (pre)}
        \label{fig:div_pre}
    \end{subfigure}
     \hfill
    \begin{subfigure}{0.32\linewidth}
        \centering
        \includegraphics[width=\linewidth]{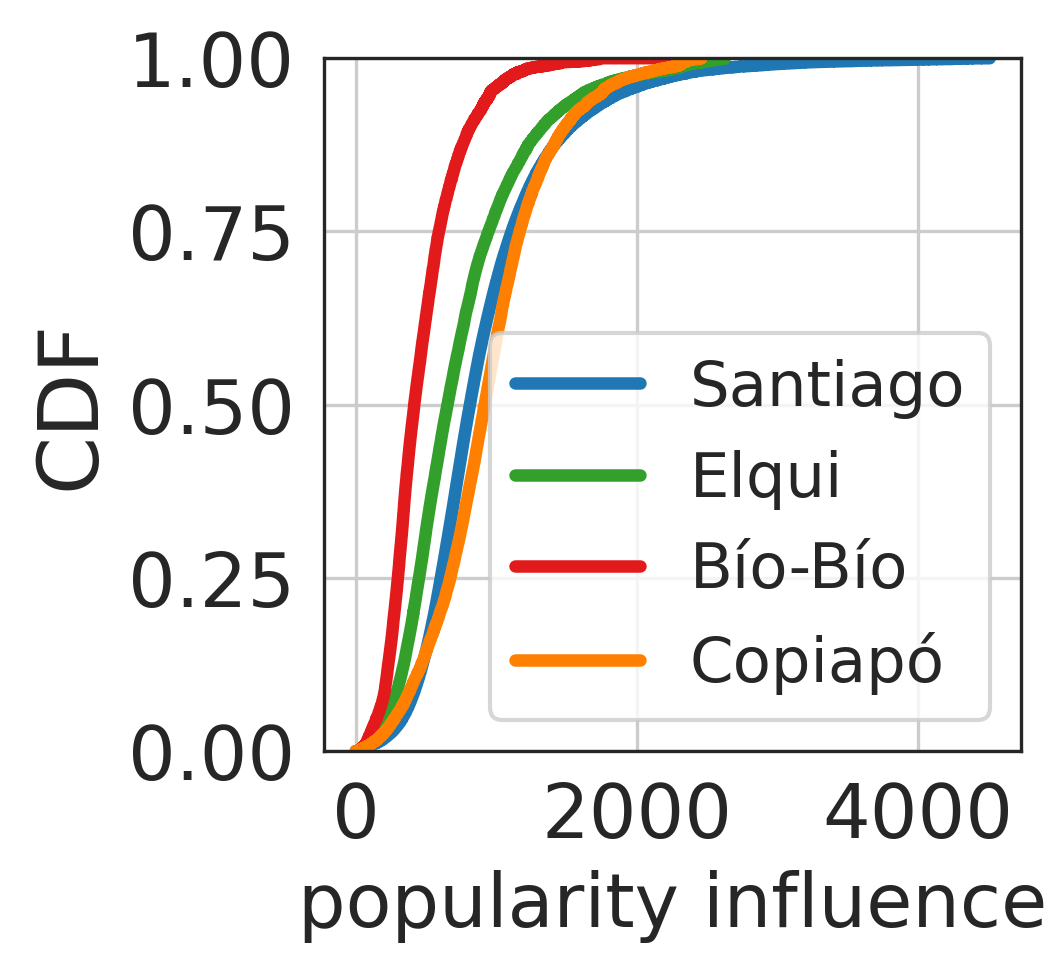}
        \caption{Popularity influence}
        \label{fig:popularity}
    \end{subfigure}
    \hfill
    \begin{subfigure}{0.32\linewidth}
        \centering
        \includegraphics[width=\linewidth]{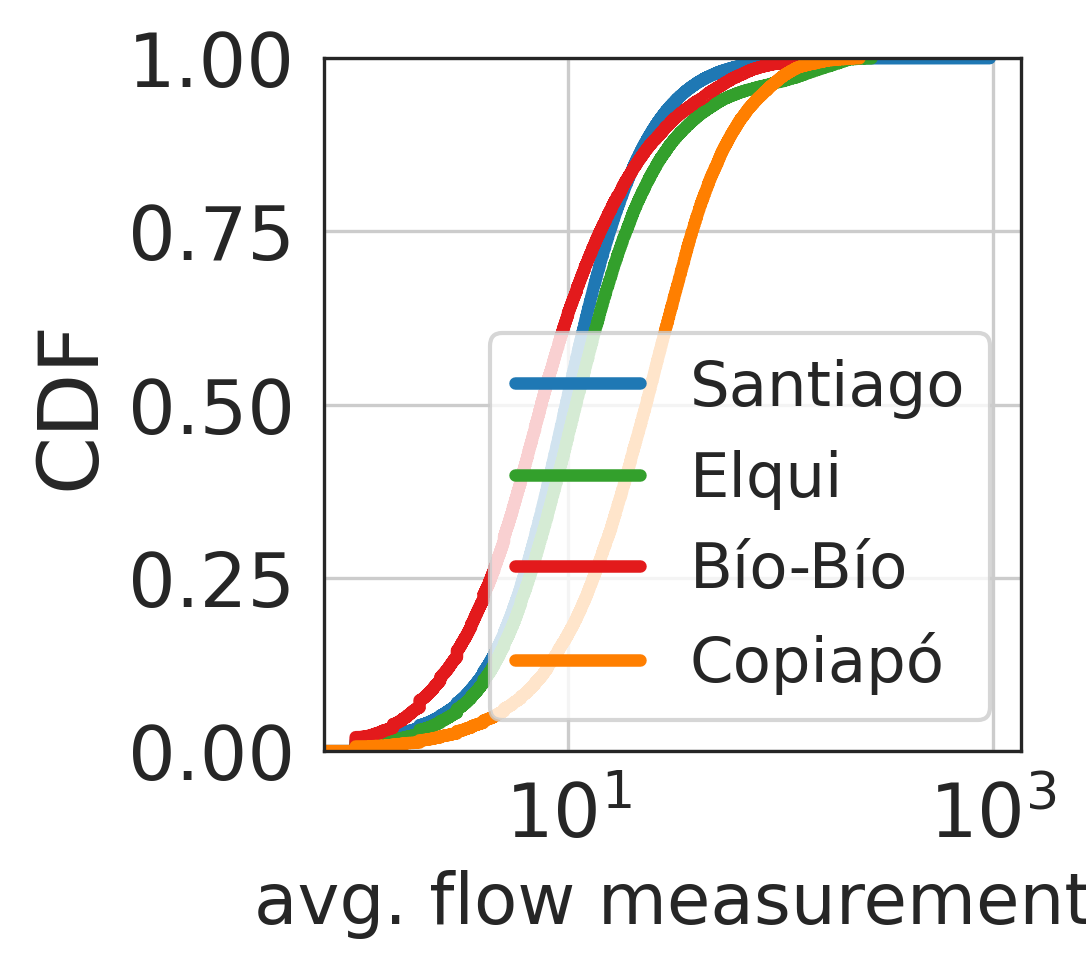}
        \caption{Flow measurement}
        \label{fig:flow}
    \end{subfigure}
    \hfill
    \begin{subfigure}{0.32\linewidth}
        \centering
        \includegraphics[width=\linewidth]{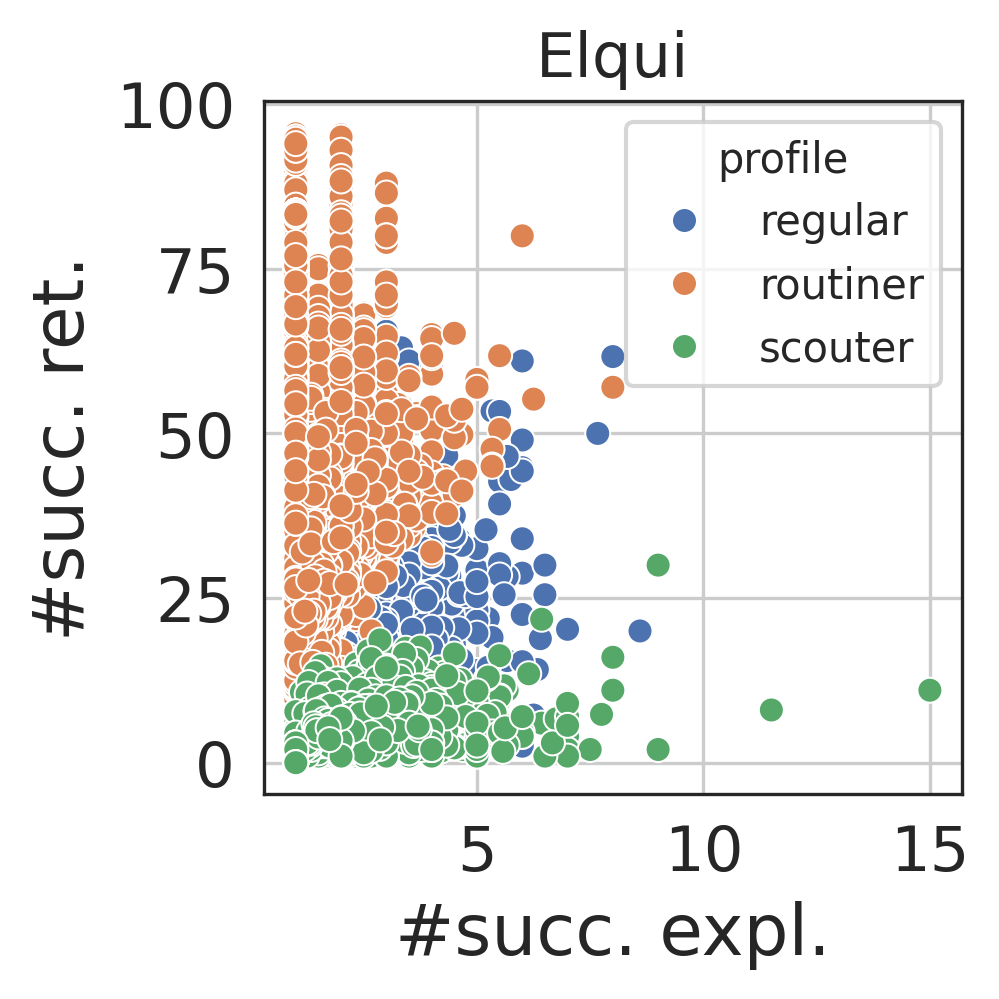}
        \caption{Mobility profiles in Elqui}
        \label{fig:mob_profiles}
    \end{subfigure}
    \caption{Users' mobility behavior features distributions}
    \label{fig:mobility-behavior}
    \vspace{-0.7cm}
\end{figure}

\vspace{0.13cm}
\noindent\textbf{Population behavior overview.} 
Fig. \ref{fig:traffic-behavior} and Fig. \ref{fig:mobility-behavior} show the distribution of mobile users' behaviors in terms of traffic and mobility features across the four studied provinces.

Fig. \ref{fig:avgnev_user} depicts the average number of events per user per day, revealing a shared behavior among provinces. 
Roughly half of the population engages in 45 to 48 events daily, a pattern characteristic of XDR datasets collecting even background data traffic. While the average traffic session volume per user is relatively consistent across provinces, uniformly ranging from 10KB to 100MB (cf. Fig. \ref{fig:avg_volume}), Fig. \ref{fig:hour_volume_norm} highlights notable differences in total hourly traffic volume normalized by the number of users in the province. Users in Santiago generate the most traffic, followed by those in Bio-Bio, while Elqui and Copiapo exhibit lower, yet similar, traffic volumes. These distributions result in approximately 40\% of users classified as \textit{LO}, 30\% as \textit{LF}, 5\% as \textit{HO}, and 25\% as \textit{HF}, with negligible variability across provinces, as shown in Fig. \ref{fig:tra_profile_norm}. 

Regarding mobility spatial features, Figs. \ref{fig:rg_unique} and \ref{fig:rg_event} show that the radius of gyration ranges between 1 km and 200 km. Notably, urbanized provinces like Santiago and Elqui exhibit lower radius of gyration values compared to Bio-Bio and Copiapo, suggesting that residents travel shorter distances due to nearby facilities in urban environments. In contrast, the higher variability observed in Elqui and Copiapo highlights the mixed nature of these provinces, which feature both urban and remote areas. Similar trends are captured in Fig. \ref{fig:avg_distance}, where the average distance traveled within 30-minute time slots remains relatively low across provinces, ranging between 100 m and 8 km. Structural mobility features, as shown in Figs. \ref{fig:rep_sta} and \ref{fig:div_pre}, reveal a high repetitiveness in user trajectories, increasing from the most urbanized province, Santiago, to the least, Copiapo. Around 80\% of the population revisits 75\% of the same locations. Stationarity levels are less pronounced but still show at least 25\% across all provinces, with higher values observed in less urbanized areas like Bio-Bio and Copiapo, despite their larger geographical sizes (cf. Table \ref{tab:data-description}). This helps explain why nearly 20\% of users exhibit almost no diversity in their movements, while the majority show a diversity index of at least 0.75. Social mobility features, depicted in Figs. \ref{fig:popularity} and \ref{fig:flow}, reflect user popularity and flow, corresponding to the overall population density in each province. However, Copiapo stands out with significantly higher values, indicating a strong clustering of the population in certain areas. Finally, Fig. \ref{fig:mob_profiles} classifies users into mobility profiles, with approximately 60\% identified as regulars, 25\% as routiners, and 15\% as explorers (cf. Fig. \ref{fig:mob_profile_norm}). Notably, Bio-Bio and Copiapo show a higher proportion of routiners, consistent with the structural mobility metrics observed.

\vspace{0.13cm}
\noindent\textbf{Interplay between User Mobility and Traffic Profiles.} 
To explore the relationship between traffic and mobility, we examine the distribution of our two traffic features across mobility profiles and our 11 mobility features across traffic profiles. Since there are only two traffic features, no filtering is needed for mobility profile analysis. However, for traffic profile analysis, we use an Ensemble Trees Classifier to predict traffic profiles (LO, LF, HO, HF) based on mobility features, selecting the most significant ones. Feature importance is assessed following \S \ref{app:feature_importance}, revealing that \textit{Rg\_unique}, \textit{diversity}, and \textit{popularity influence} are the most impactful spatial, structural, and social mobility features, consistent across all provinces (cf. Fig. \ref{fig:fi_mob_for_traffi}).

\begin{figure*}
\vspace{-0.4cm}
    \centering
    \begin{subfigure}{0.16\linewidth}
        \centering
        \includegraphics[width=\linewidth, height=3cm, keepaspectratio]{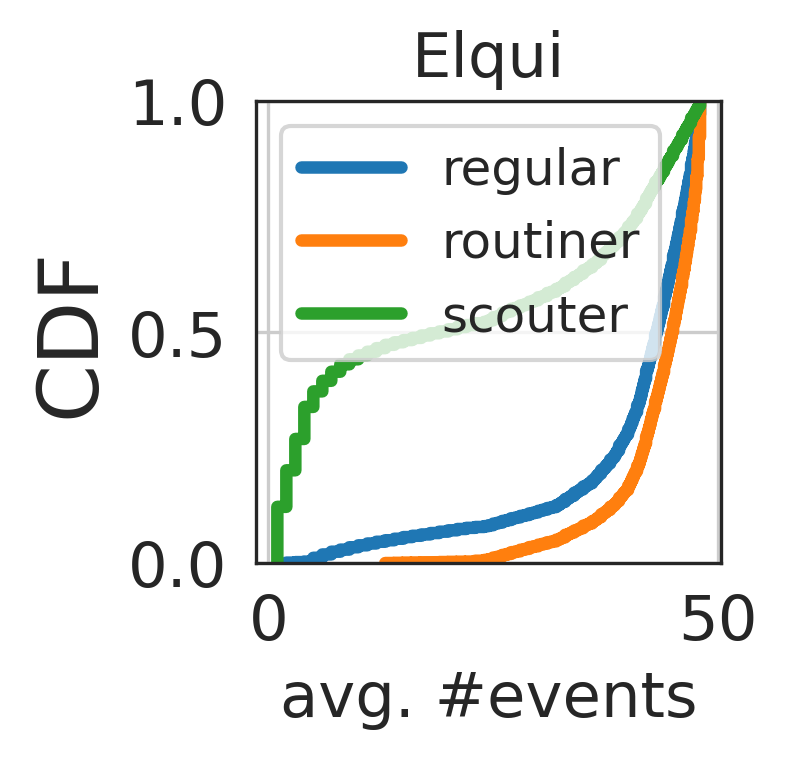}
        \caption{Avg. \#events}
        \label{fig:avg_nev_profile}
    \end{subfigure}
    \hfill
     \begin{subfigure}{0.21\linewidth}
        \centering
        \includegraphics[width=\columnwidth, height=3cm, keepaspectratio]{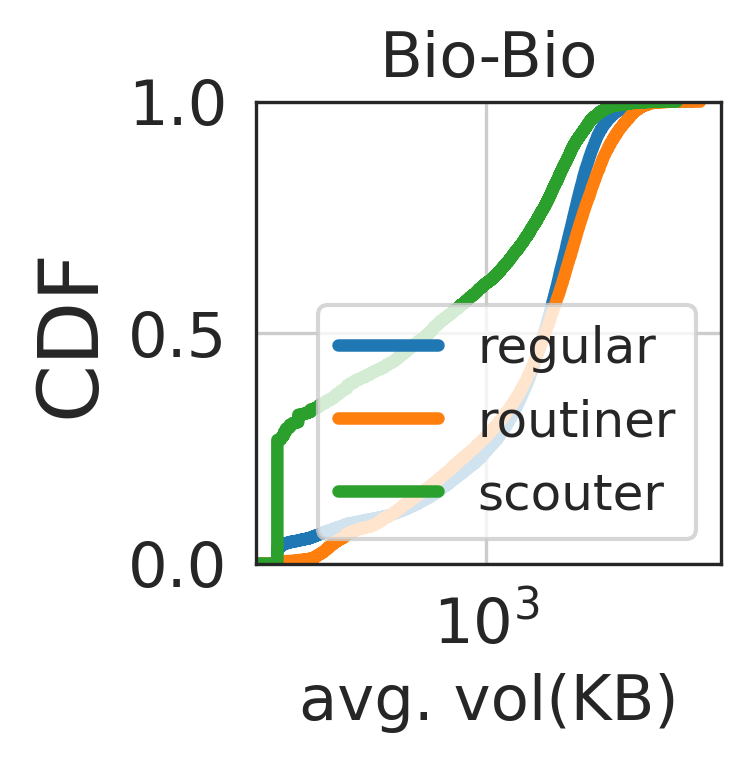}
        \caption{Avg. volume}
        \label{fig:hour_vol_profile}
    \end{subfigure}
    \hfill
     \begin{subfigure}{0.16\linewidth}
        \centering
        \includegraphics[width=\linewidth, height=3cm, keepaspectratio]{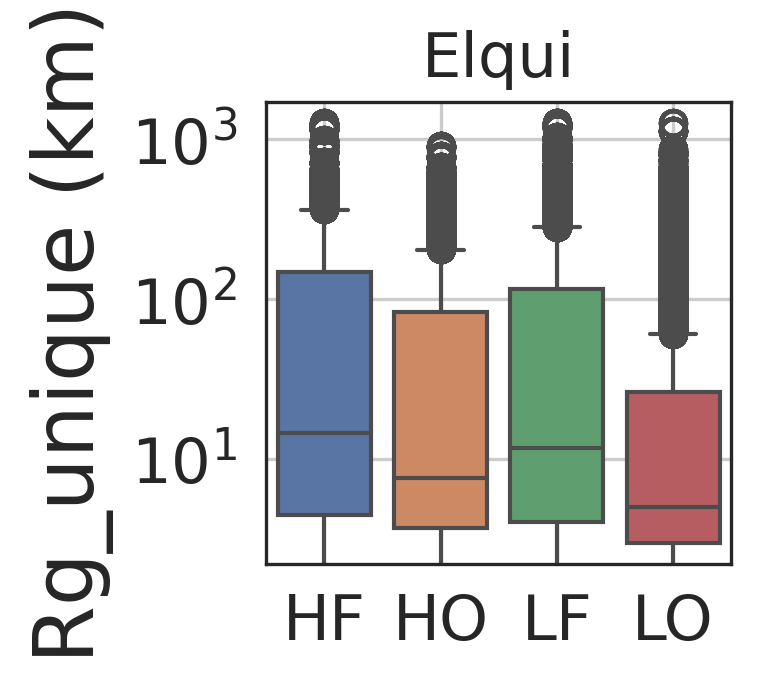}
        \caption{Rg unique}
        \label{fig:rg_unique_profile}
    \end{subfigure}
    \hfill
    \begin{subfigure}{0.16\linewidth}
        \centering
        \includegraphics[width=\linewidth, height=3cm, keepaspectratio]{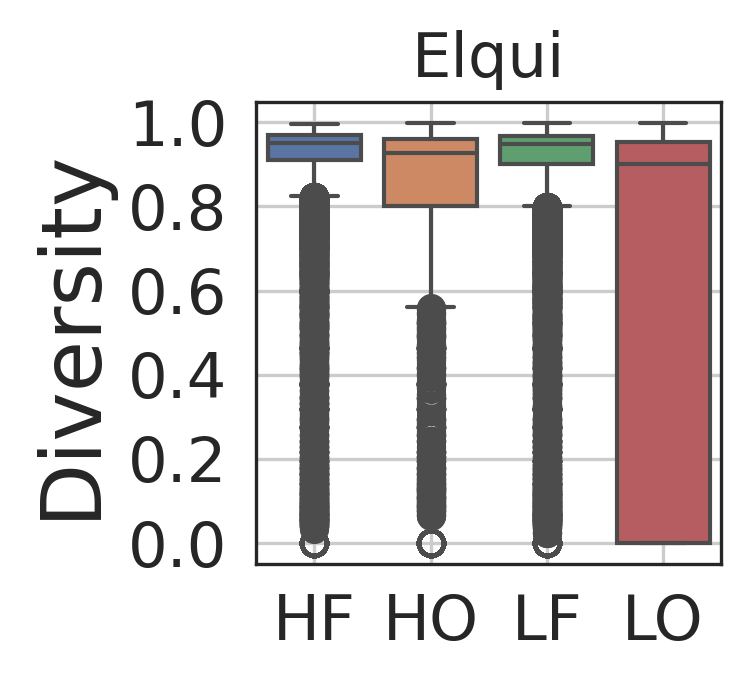}
        \caption{Diversity}
        \label{fig:div_profile}
    \end{subfigure}
    \hfill
    \begin{subfigure}{0.16\linewidth}
        \centering
        \includegraphics[width=\linewidth, height=3cm, keepaspectratio]{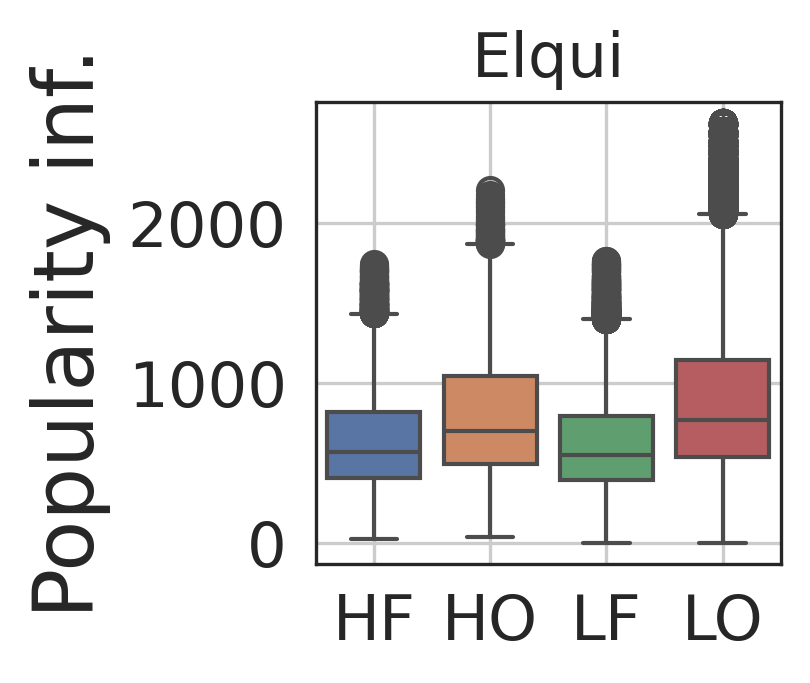}
        \caption{Popularity influence}
        \label{fig:pop_infl_profile}
    \end{subfigure}
     \caption{(a)-(b) Traffic features by mobility profile; (c)-(e) Mobility features distribution by traffic profile.}
    \label{fig:tra_mob_interplay}
    \vspace{-0.2cm}
\end{figure*}

The analyses computed in Elqui, shown in Fig. \ref{fig:tra_mob_interplay}, enhance our understanding of user behavior yielding the following insights:
\greybox{\textbf{Takeaway 1 (Figs \ref{fig:avg_nev_profile}, \ref{fig:hour_vol_profile}):} \textit{Scouters, or extreme explorers, generate traffic significantly less frequently and at lower volumes compared to regulars and routiners. While the average traffic volume for regulars and routiners is quite similar, it is noteworthy that regulars engage in traffic generation less often than routiners. This suggests that more exploratory users tend to participate in activities that yield less traffic and occur less frequently. In contrast, routiners, being more stationary, have greater opportunities for high-traffic activities.}

\textbf{Takeaway 2 (Figs \ref{fig:rg_unique_profile}, \ref{fig:div_profile}, \ref{fig:pop_infl_profile}):} \textit{Users with an occasional traffic profile (LO and FO) demonstrate a lower radius of gyration (Rg) compared to those who engage more frequently in traffic. This aligns with preliminary insights from the literature~\cite{Paul:2011, Alipour:2018}, indicating that spatial extent correlates first with traffic frequency and then with volume. Additionally, users with occasional traffic show lower trajectory diversity, while those with higher frequency traffic profiles (LF and HF) possess greater diversity in their movements. This trend reflects that increased traffic engagement relates to more varied paths. Finally, the popularity influence is higher for occasional users, suggesting that they are drawn to popular locations that may not require frequent or heavy traffic generation, but likely engage them in activities of a different nature, such as shopping or socializing.}}


\section{Mobile User Behavior modeling}
\label{sec:contrib2}
The previous characterizations, in line with approaches in the literature, represent each user as a point in an n-dimensional vector space, defined by a fixed set of traffic and mobility metrics. While this method allows for preliminary dependency exploration, it imposes a rigid, coarse-grained characterization of users. Assigning static profiles (e.g., Heavy or Light) to users tends to overlook the heterogeneous and evolving nature of their behaviors, and it fails to capture more granular dependencies present in XDR data.

We extend the analysis by proposing a dynamic and flexible user modeling approach that captures dependencies over time between mobility and traffic behaviors. This allows for a finer understanding of user interactions, leading to more accurate and context-agnostic insights.
The details of this modeling are formalized in \S \ref{subsec:formalization}, and its analytical implications are discussed in \S \ref{subsec:granular_analysis}.

\subsection{Formal definition}
\label{subsec:formalization}
\textit{We model each user as a sequence of evolving mobility and traffic behaviors over time, capturing the dynamic nature of their interactions.} Specifically, for each time slot $t_i$, we account for traffic and mobility features characteristic of the user’s behavior during that time step.
Therefore, $u = (b^u_1, b^u_2, \dots, b^u_n)$ with $b^u_i = \left( t_i, b^u_{\text{tra-i}}, b^u_{\text{mob-i}} \right)$ being a behavior tuple charactering the user at time step $t_i$. $b^u_{\text{tra-i}}$ represents the step-level traffic behavior while $b^u_{\text{mob-i}}$ captures the corresponding mobility features. These step-level features are derived from the global features as defined in Table \ref{tab:features_desc} and discussed below.

\vspace{0.13cm}
\noindent\textbf{Traffic behavior.} 
We categorize each step-level volume $v_i^u$ as \textit{light} (\( l \)), \textit{medium} (\( m \)), or \textit{heavy} (\( h \)), based on the quantiles of the global session volume distribution. This allows us to track the user’s traffic evolution as a sequence \( (trc_1, trc_2, \dots, trc_n) \), where $trc_i \in \{l, m, h\}$. Traffic frequency is inherently captured by the length of the sequence, therefore, $b^u_{\text{tra-i}} = (trc^u_i)$ reflects only the volume.

\vspace{0.13cm}
\noindent\textbf{Mobility behavior:} For modeling mobility at each step, we use selected spatial, structural, and social metrics. The \textbf{spatial features} include the \textit{traveled distance} $d_i$ between consecutive locations. Similar to traffic, we classify $d_i$ into three categories: \textit{close, medium,} or \textit{far}, based on the quantiles of the global step-level distances distribution.
The \textit{radius of gyration} is excluded from step-based modeling as it requires the full trajectory to compute the center of mass.
For \textbf{structural features}, we track the \textit{repetitiveness} of each step as a boolean state being either a return to a previous location (i.e., 1) or exploration of a new one (i.e., 0), and \textit{stationarity} as a boolean state indicating if the user remains at the same location (\( l_i = l_{i+1} \)). We also assess the change in trajectory diversity at each step through \textit{delta-diversity} (\( \delta div_i \)) while avoiding a separate predictability measure, as it is strongly influenced by diversity.
For \textbf{social features}, we capture the \textit{zone popularity} of the locations visited by the user at each step \( p(l_i, t_i) \) and measure the flow of users $flow(l_i, l_{i+1}, t_i)$ following the same movement path at the same time step. 

Consequently, the mobility behavior at each step is represented as \( b^u_{\text{mob-i}} = \left( disc_i, rep_i, sta_i, \delta \text{div}_i, p(l_i, t_i), flow(l_i, l_{i+1}, t_i) \right) \).


\subsection{Granular dependency Analysis}
\label{subsec:granular_analysis}

The current analyses examine the dependency between step-level traffic behaviors and individual mobility features, identifying mobility indicators most affected by traffic patterns. These insights are then used to refine the user modeling.

\vspace{0.13cm}
\noindent \textbf{Step-level dependency analysis.}  
We analyze the dependency between traffic behavior and mobility features at each time step. Across all provinces, users tend to generate light traffic events most frequently, followed by medium and heavy traffic (cf. Fig. \ref{fig:step_entire_distrib} in the Appendix). Interestingly, only heavy traffic volumes exhibit a clear daily pattern, peaking during the day and dropping at night, as shown in Fig. \ref{fig:step_hourly_distrib}. In contrast, light traffic is more prominent at night, reflecting background activity when users are generally inactive, and declines during the day as heavier traffic takes over.

\begin{figure*}
\vspace{-0.2cm}
    \centering
    \begin{subfigure}{0.16\linewidth}
        \centering
        \includegraphics[width=\linewidth]{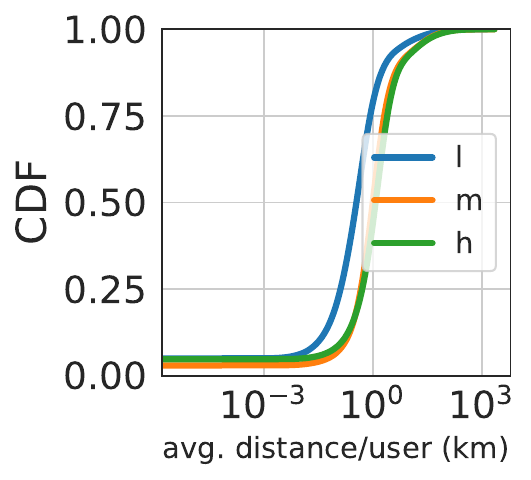}
        \caption{Traveled distance}
        \label{fig:avg_dis_step}
    \end{subfigure}
    \hfill
    \begin{subfigure}{0.16\linewidth}
        \centering
        \includegraphics[width=\linewidth]{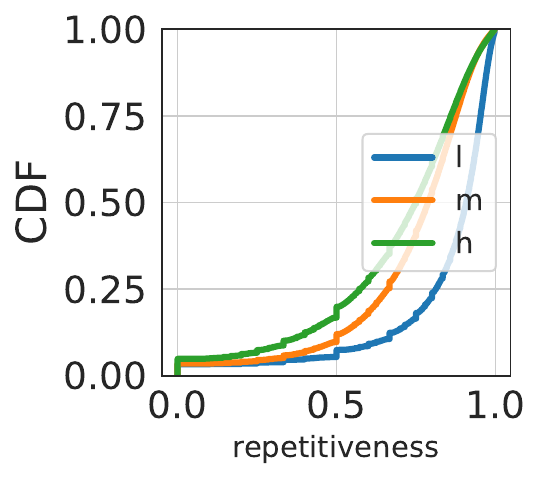}
        \caption{Repetitiveness}
        \label{fig:rep_step}
    \end{subfigure}
     \hfill
    \begin{subfigure}{0.16\linewidth}
        \centering
        \includegraphics[width=\linewidth]{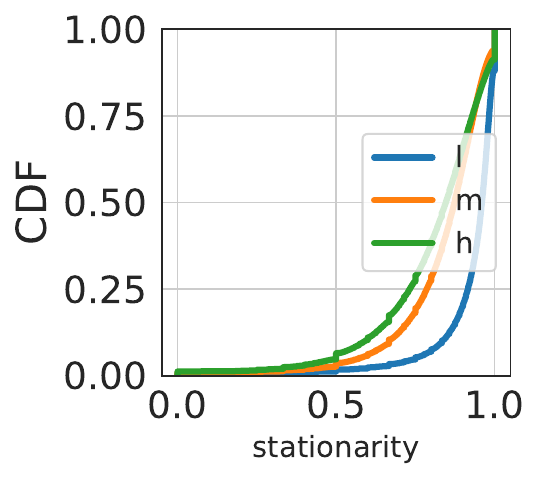}
        \caption{Stationarity}
        \label{fig:sta_step}
    \end{subfigure}
     \hfill
     \begin{subfigure}{0.16\linewidth}
        \centering
        \includegraphics[width=\linewidth]{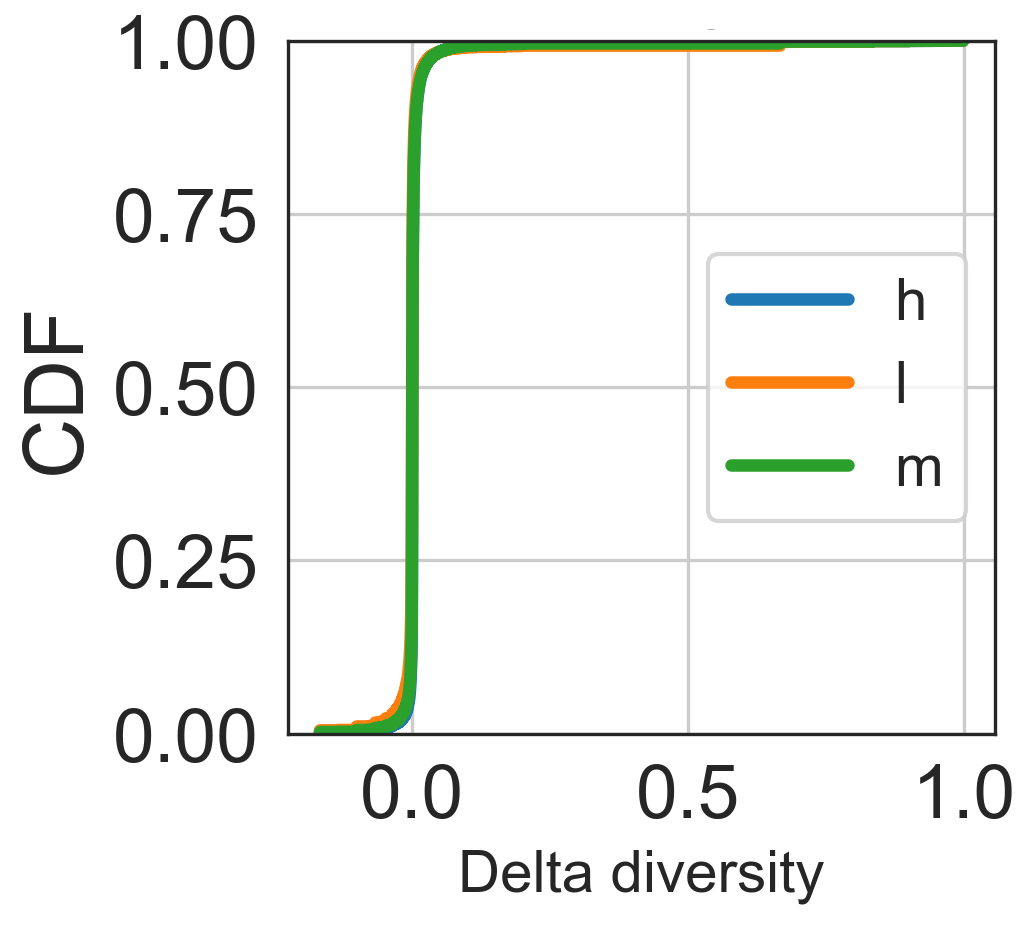}
        \caption{Delta diversity}
        \label{fig:div_step}
    \end{subfigure}
     \hfill
    \begin{subfigure}{0.16\linewidth}
        \centering
        \includegraphics[width=\linewidth]{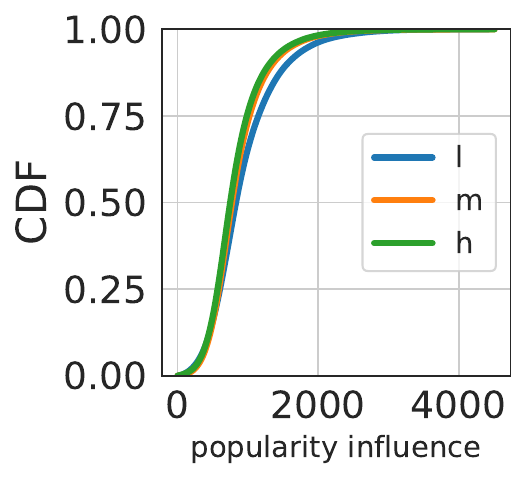}
        \caption{Popularity influence}
        \label{fig:pop_step}
    \end{subfigure}
     \hfill
    \begin{subfigure}{0.16\linewidth}
        \centering
        \includegraphics[width=\linewidth]{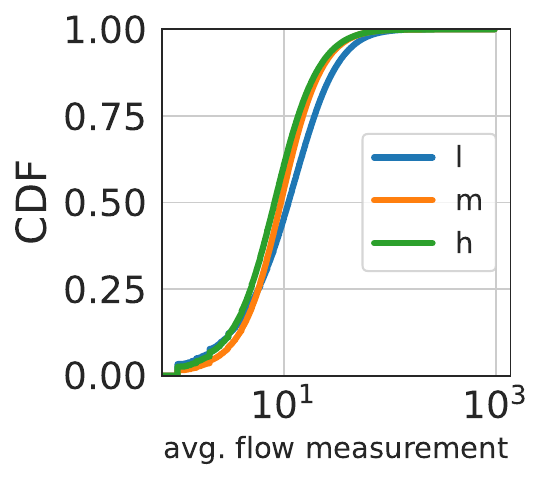}
        \caption{Flow measurement}
        \label{fig:flow_step}
    \end{subfigure}
    \caption{Dependency between in-step traffic and mobility features of users in Santiago}
    \label{fig:in_step_correlation}
    \vspace{-0.2cm}
\end{figure*}

To establish dependency, we examine how different mobility features distribute across traffic categories (\textit{light}, \textit{medium}, \textit{heavy}). Significant differences in the distributions suggest strong dependencies between certain mobility features and traffic patterns. Our results, illustrated in Fig. \ref{fig:in_step_correlation}, highlight three mobility features strongly influenced by traffic category: \textit{traveled distance} (cf. Fig. \ref{fig:avg_dis_step}), \textit{repetitiveness} (cf. Fig. \ref{fig:rep_step}), and \textit{stationarity} (cf. Fig. \ref{fig:sta_step}). In contrast, mobility features like \textit{popularity} (cf. Fig. \ref{fig:pop_step}) and \textit{flow} (cf. Fig. \ref{fig:flow_step}) show only moderate dependency with traffic, and \textit{delta diversity} (cf. Fig. \ref{fig:div_step}) appears to be unaffected. Key insights emerge from these findings: 
\vspace{0.2cm}
\greybox{
\textbf{Takeaway 3:} \textit{Users tend to make shorter trips during light traffic periods, suggesting that light traffic is often generated by background activities when the user is less active, naturally reducing mobility. In contrast, users involved in medium or heavy traffic are likely traveling longer distances, as part of their daily routines. Additionally, repetitive and stationary movement patterns are more common during light traffic and decrease as traffic intensity rises, implying that during heavy traffic, users are likely engaging in more varied or spontaneous activities that disrupt routine behaviors. This supports the idea that predictable, stationary actions are linked to lighter traffic, while more dynamic behaviors align with higher traffic volumes.}
}

\vspace{0.13cm} \noindent \textbf{Refining user modeling.}
Building on the step-level dependency analysis, the user modeling framework is refined by selectively retaining only the mobility features that exhibit a strong dependency with traffic behavior. Specifically, the mobility features of \textit{traveled distance}, \textit{repetitiveness}, and \textit{stationarity} emerge as statistically relevant to traffic dynamics. By focusing on these key features, the model's complexity is reduced, improving interpretability without sacrificing accuracy. This refined representation not only streamlines the analysis but also ensures that the most impactful mobility traits are used to explain user behavior. As a result, it enhances both the clarity and the computational efficiency of the user modeling approach, forming the basis for the remainder of the paper.

\section{Behavior Inference}
\label{sec:contrib3}
This section builds upon the proposed user modeling approach to develop a methodology for inferring a mobile user's traffic behavior from their mobility patterns, and vice versa.
Our methodology is grounded in two key principles:

\begin{itemize}[leftmargin=*]
    \item \textit{Realism:} The refined statistical dependency between traffic behavior \(b_{\text{tra-i}}\) and mobility behavior \(b_{\text{mob-i}}\) enables to assess whether a step-level behavior is realistic.
    For instance, as higher traffic volumes (\textit{light} \(l\) to \textit{heavy} \(h\)) are positively correlated with increased \textit{traveled distance} (\textit{close} to \textit{far}), the state \(lc\) (light traffic, close distance) as realistic, whereas \(lf\) (light traffic, far distance) is less plausible. 
    This enables us to evaluate the statistical coherence of step-level behaviors and transitions between them.
    \item \textit{Complexity:} For each user, the sequence of realistic behaviors across time forms an increasingly unique signature. As the length of this sequence grows, it becomes more challenging to find another user with the exact same sequence of traffic behaviors $(b_{\text{tra-0}},$ $ \dots, b_{\text{tra-n}})$ or mobility behaviors $(b_{\text{mob-0}},$ $ \dots, b_{\text{mob-n}})$. The longer the sequence, the greater the specificity, and thus the more distinctive the user's behavior becomes. 
\end{itemize}
By integrating these principles, \textit{we introduce a model that captures the likelihood of a user’s sequence by aligning traffic behaviors with corresponding mobility behaviors.} \S \ref{subsec:model_definition} formally defines the model and its likelihood metric, while \S \ref{subsec:inference_use_cases} presents inference use cases and performance evaluation on our dataset.

\subsection{Model definition}
\label{subsec:model_definition}
Given a population \( \mathcal{U} = \{ u_k \}_{k=1}^{P} \), where each user \( u_k \) is represented as a sequence of temporal states 
\[u_k = (t_0^k b_{\text{tra-0}}^k b_{\text{mob-0}}^k, \dots, t_n^k b_{\text{tra-n}}^k b_{\text{mob-n}}^k)\]
we define the inference model as a Markov model \( M = (S, P) \) where:
\begin{itemize}[leftmargin=*]
    \item Each state \( s \in S \) corresponds to a unique combination of behavior defined as \( s_i = (t_i b_{\text{tra-i}} b_{\text{mob-i}}) \).
    \item The transition matrix \( P \) defines the probabilities of transitioning between states. Each element \( p_t = p_{ij} \) represents the probability of transitioning from state \( s_i \) to state \( s_j \) across of the population. This means that \( p_t = p_{ij} > 0\) if there is at least one user \( u_k \) in the population and state \( l \) such that \( (s_i, s_j) = (s_l^k, s_{l+1}^k) \).
\end{itemize}

\textit{The proposed inference model allows to compute the likelihood of matching a sequence of traffic behaviors to a corresponding sequence of mobility behaviors through the following algorithm:}
\begin{enumerate}[leftmargin=*]
    \item Given a sequence of traffic behaviors \( (b_{\text{tra-0}}, \dots, b_{\text{tra-n}}) \) and mobility behaviors \( (b_{\text{mob-0}}, \dots, b_{\text{mob-n}}) \), we extract the corresponding sequence consisting of states \( s_0, \ldots, s_n \), where \( s_i = (t_i b_{\text{tra-i}} b_{\text{mob-i}}) \).
    
    \item Next, from the obtained sequence, we calculate several metrics:
    \begin{itemize}[leftmargin=*]
        \item \( r_t = \frac{n_{\text{valid}}}{n_{\text{total}}} \): the rate of unique valid transitions in the sequence, i.e., with a non-zero transition probability \( p_t > 0 \).
        \item \( p_{\text{all-t}} = \left( \prod_{j=1}^{n_{\text{valid}}} p_{t_j} \right)^{\frac{1}{n_{\text{valid}}}} \): the n-th root of the product of probabilities for the valid transitions in the sequence, representing the geometric mean likelihood.

    \end{itemize}
    
    The overall sequence likelihood is: $\pi = \alpha \cdot r_t + (1 - \alpha) \cdot p_{\text{all-t}}$
    \\where $\alpha$ is an empirical parameter balancing the contributions of the rate and the likelihood of valid transitions in the process.
\end{enumerate}

\begin{figure}[h!]
    \centering
    \includegraphics[width=\linewidth]{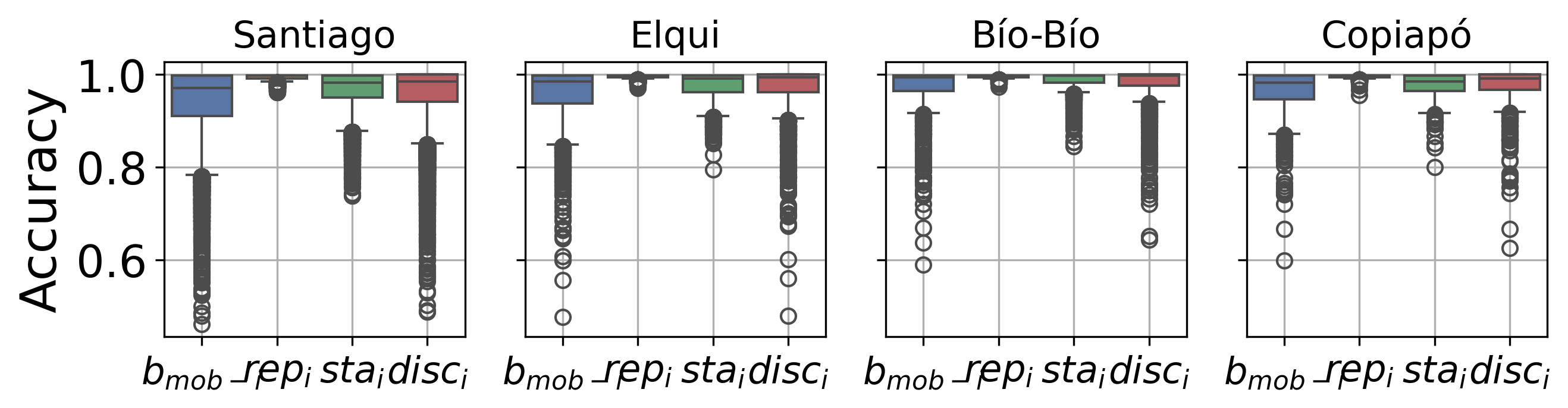}
    \caption{Accuracy of inferring mobility from traffic behavior}
    \label{fig:mobtotra_usecase1}
    \vspace{-0.1cm}
\end{figure}

\begin{table}[h!]
\vspace{-0.2cm}
\centering
\caption{Accuracy of reconstructing actual users}
\label{tab:inference3}
\resizebox{\columnwidth}{!}{%
\begin{tabular}{lllllllll}
\multicolumn{1}{c}{} &
  \multicolumn{1}{c}{\rotatebox{90}{\textit{\textbf{\begin{tabular}[c]{@{}c@{}}Santiago\\ distrib\end{tabular}}}}} &
  \multicolumn{1}{c}{\rotatebox{90}{\textit{\textbf{\begin{tabular}[c]{@{}c@{}}Santiago\\ routiner\end{tabular}}}}} &
  \multicolumn{1}{c}{\rotatebox{90}{\textit{\textbf{\begin{tabular}[c]{@{}c@{}}Santiago\\ regular\end{tabular}}}}} &
  \multicolumn{1}{c}{\rotatebox{90}{\textit{\textbf{\begin{tabular}[c]{@{}c@{}}Santiago\\ scouter\end{tabular}}}}} &
  \multicolumn{1}{c}{\rotatebox{90}{\textit{\textbf{\begin{tabular}[c]{@{}c@{}}Santiago\\ uniform\end{tabular}}}}} &
  \multicolumn{1}{c}{\rotatebox{90}{\textit{\textbf{Elqui}}}} &
  \multicolumn{1}{c}{\rotatebox{90}{\textit{\textbf{Bio-Bio}}}} &
  \multicolumn{1}{c}{\rotatebox{90}{\textit{\textbf{Copiapo}}}} \\  \cline{2-9} 
\multicolumn{1}{l|}{\textbf{Top-1}} &
  \multicolumn{1}{l|}{66\%} &
  \multicolumn{1}{l|}{66\%} &
  \multicolumn{1}{l|}{65\%} &
  \multicolumn{1}{l|}{64\%} &
  \multicolumn{1}{l|}{64\%} &
  \multicolumn{1}{l|}{51\%} &
  \multicolumn{1}{l|}{46\%} &
  \multicolumn{1}{l|}{63\%} \\ \cline{2-9} 
\multicolumn{1}{l|}{\textbf{Top-5\%}} &
  \multicolumn{1}{l|}{70\%} &
  \multicolumn{1}{l|}{70\%} &
  \multicolumn{1}{l|}{70\%} &
  \multicolumn{1}{l|}{69\%} &
  \multicolumn{1}{l|}{68\%} &
  \multicolumn{1}{l|}{53\%} &
  \multicolumn{1}{l|}{47\%} &
  \multicolumn{1}{l|}{65\%} \\ \cline{2-9} 
\multicolumn{1}{l|}{\textbf{Top-10\%}} &
  \multicolumn{1}{l|}{71\%} &
  \multicolumn{1}{l|}{70\%} &
  \multicolumn{1}{l|}{70\%} &
  \multicolumn{1}{l|}{70\%} &
  \multicolumn{1}{l|}{68\%} &
  \multicolumn{1}{l|}{53\%} &
  \multicolumn{1}{l|}{50\%} &
  \multicolumn{1}{l|}{73\%} \\ \cline{2-9} 
\multicolumn{1}{l|}{\textbf{Top-15\%}} &
  \multicolumn{1}{l|}{71\%} &
  \multicolumn{1}{l|}{71\%} &
  \multicolumn{1}{l|}{71\%} &
  \multicolumn{1}{l|}{71\%} &
  \multicolumn{1}{l|}{68\%} &
  \multicolumn{1}{l|}{53\%} &
  \multicolumn{1}{l|}{50\%} &
  \multicolumn{1}{l|}{73\%} \\ \cline{2-9} 
\multicolumn{1}{l|}{\textbf{Top-20\%}} &
  \multicolumn{1}{l|}{72\%} &
  \multicolumn{1}{l|}{72\%} &
  \multicolumn{1}{l|}{71\%} &
  \multicolumn{1}{l|}{71\%} &
  \multicolumn{1}{l|}{70\%} &
  \multicolumn{1}{l|}{55\%} &
  \multicolumn{1}{l|}{90\%} &
  \multicolumn{1}{l|}{95\%} \\ \cline{2-9} 
\end{tabular}%
}
\vspace{-0.6cm}
\end{table}

\begin{figure*}[h!]
    \centering
    \begin{subfigure}{0.35\linewidth}
        \centering
        \includegraphics[width=\linewidth]{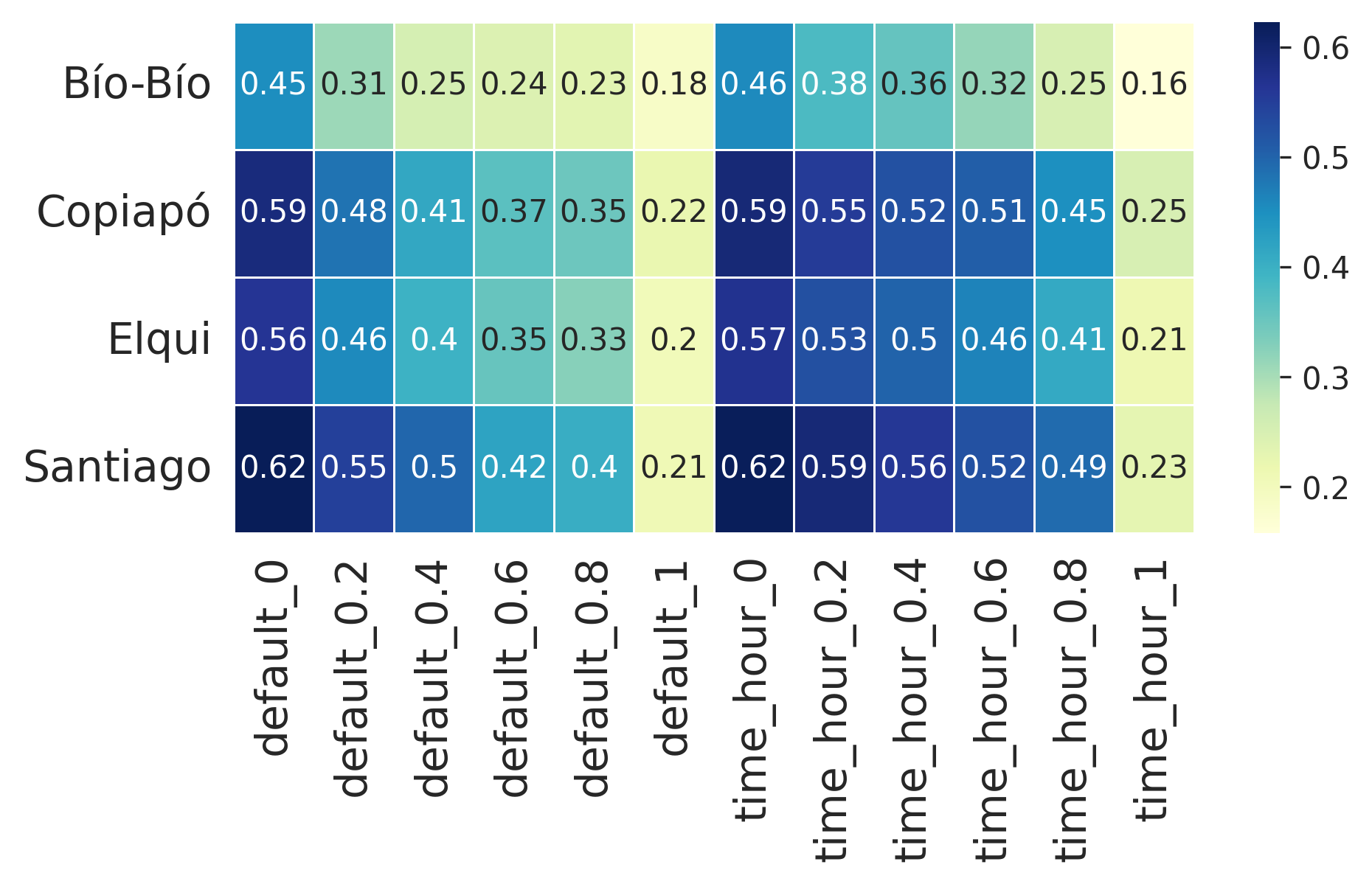}
        \caption{Intra-province w.r.t. the model version and $\alpha$}
        \label{fig:hdis-intra}
    \end{subfigure}
    \hfill
     \begin{subfigure}{0.23\linewidth}
        \centering
        \includegraphics[width=\linewidth]{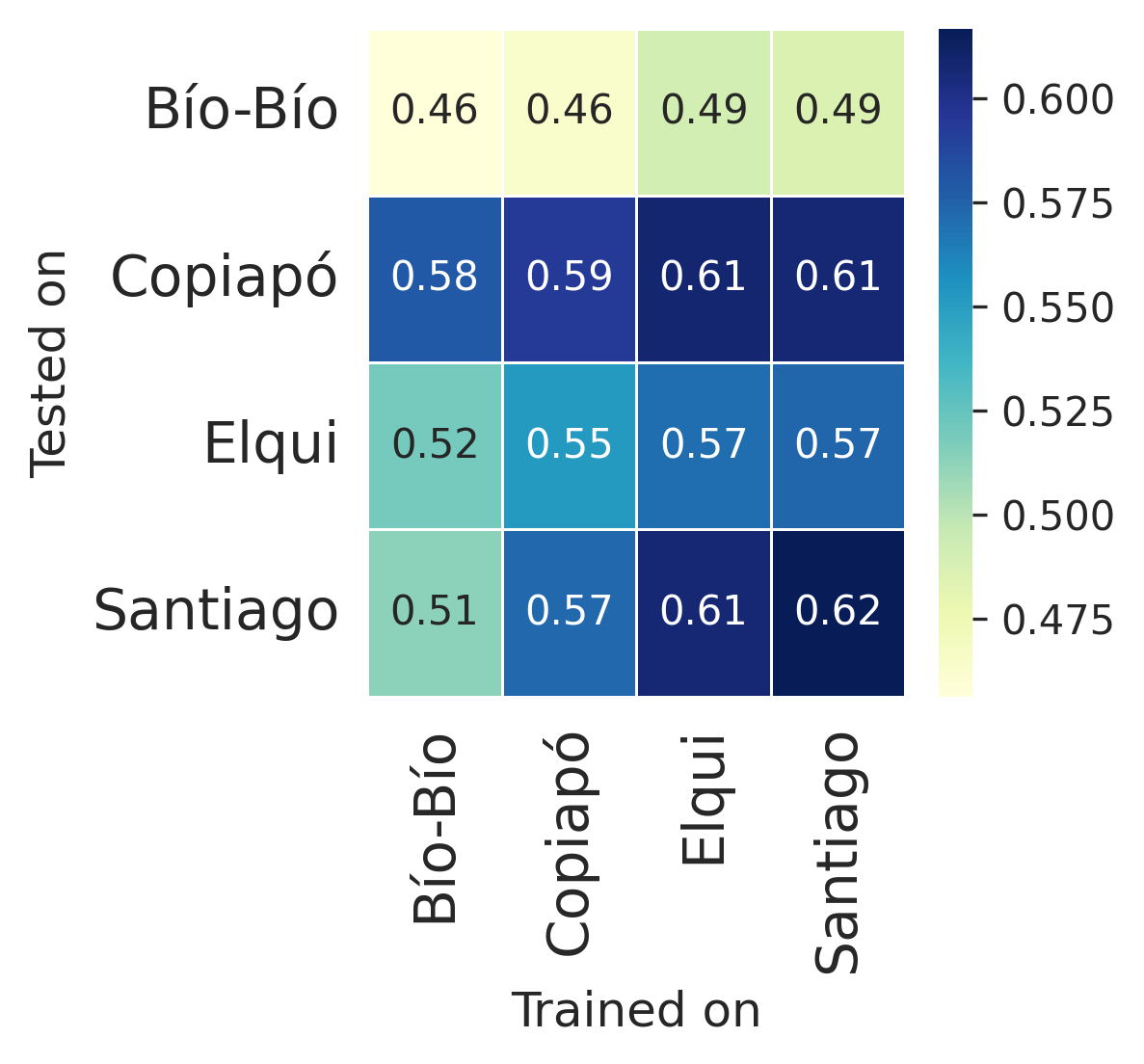}
        \caption{Inter-province with \textit{time-hour} version and $\alpha=0$}
        \label{fig:hdis-inter}
    \end{subfigure}
    \hfill
    \begin{subfigure}{0.2\linewidth}
        \centering
        \includegraphics[width=\linewidth]{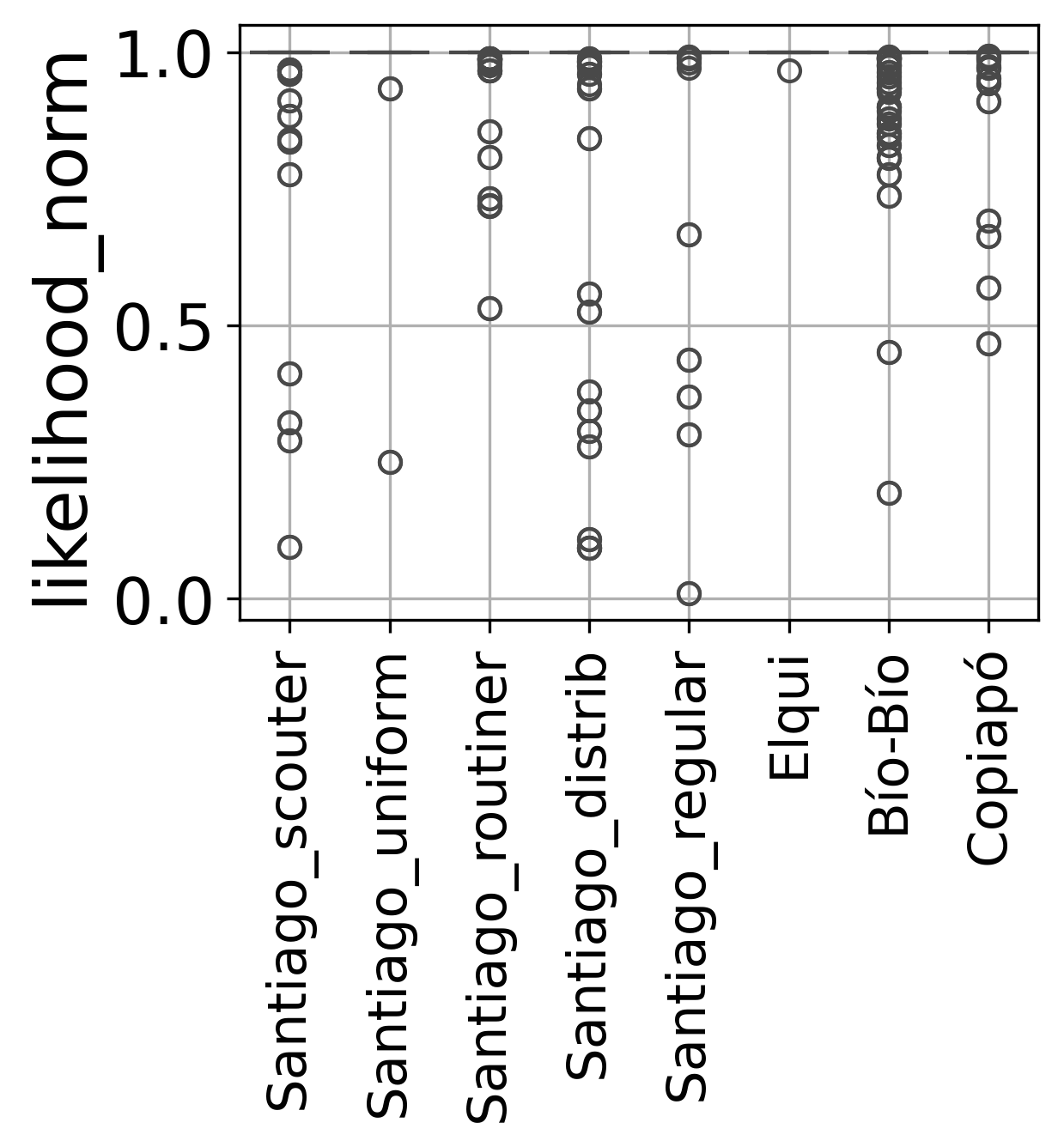}
        \caption{Min-max-normalized ground-truth likelihood}
        \label{fig:max_likelihood}
    \end{subfigure}
     \hfill
    \begin{subfigure}{0.2\linewidth}
        \centering
        \includegraphics[width=\linewidth]{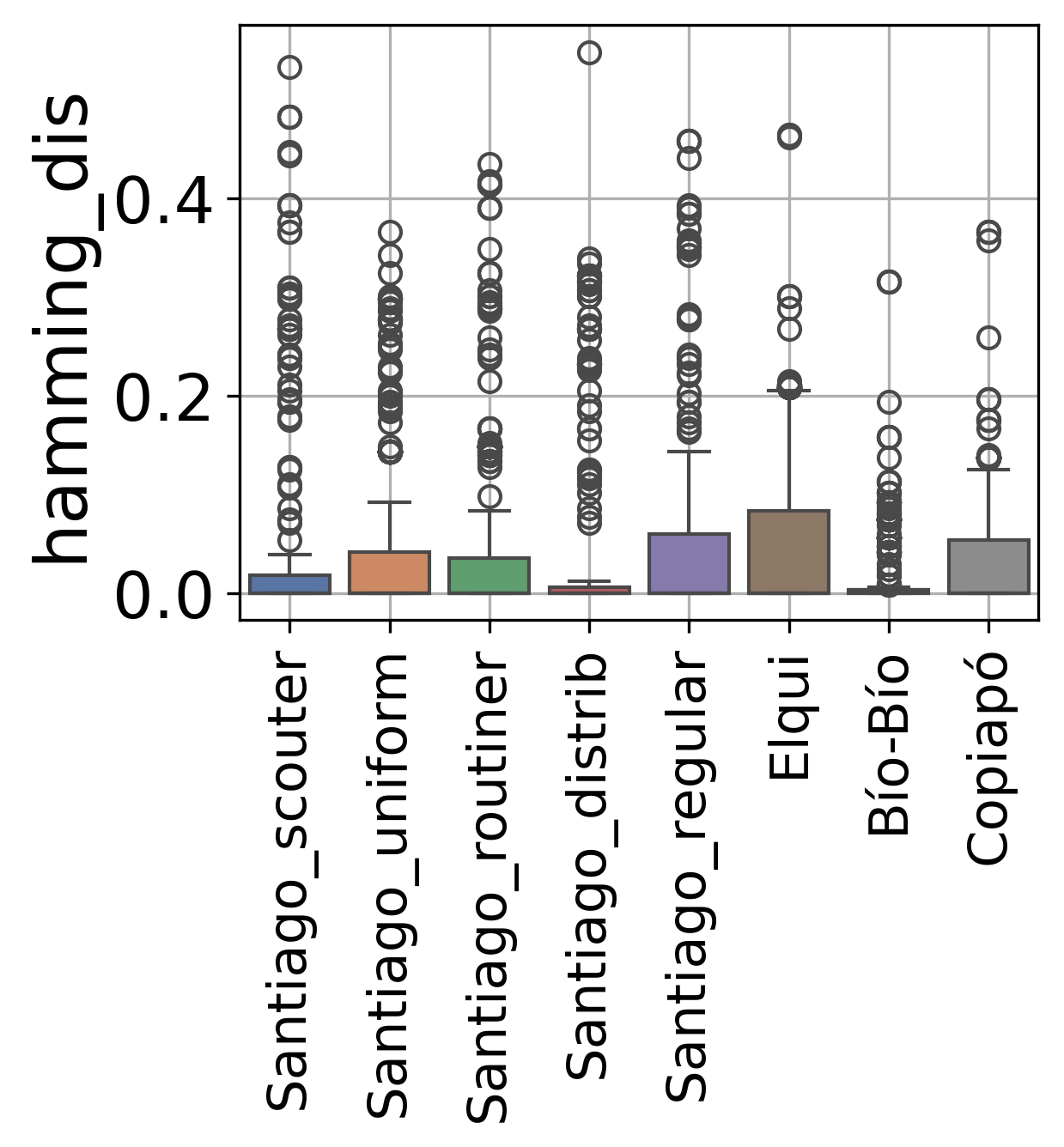}
        \caption{Normalized hamming distance to the ground truth}
        \label{fig:max_hamming_dis}
    \end{subfigure}
    \caption{Inference performance: (a)-(b) Hellinger distance between regular and shuffled likelihood distributions, (c)-(d) Traffic-Mobility matching parameters}
    \label{fig:daily-distrib}
    \vspace{-0.3cm}
\end{figure*}

\subsection{Inference use cases}
\label{subsec:inference_use_cases}

Utilizing the trained inference model, we explore three use cases:
\begin{enumerate}[leftmargin=*]
    \item \textbf{Predicting behavior sequences:} We evaluate the model's ability to predict a user's sequence of mobility behavior from their traffic behavior, and vice versa. By repeatedly sampling from the trained Markov model, we generate sequences that match the user's known behavioral states (either traffic or mobility) and their respective transition probabilities. This process allows us to measure the prediction accuracy of inferring traffic behavior from mobility patterns and vice versa, providing insights into how well the model captures these dynamics.
    \item \textbf{Distinguishing realistic versus non-realistic users:} The second use case examines the model's ability to differentiate between realistic and non-realistic user behaviors. A realistic user features correctly matched traffic and mobility sequences, while a non-realistic one is generated by shuffling and mismatching these sequences across different users. To test this, we create two datasets: one with \textit{shuffled} traffic and mobility sequences representing non-realistic users, and the \textit{regular} one with actual matches. The trained model then computes the likelihood distribution for both datasets, which is compared to assess how effectively it identifies realistic patterns.
    \item \textbf{Reconstructing integrated datasets:} In this use case, we evaluate the model’s ability to reconstruct integrated datasets by realistically matching traffic data to the corresponding mobility data. We first separate the datasets into traffic and mobility sequences, then match each traffic sequence to the most likely corresponding mobility sequence based on the highest likelihood scores. The model’s performance is assessed by (i) the accuracy of matching users’ actual behavior and (ii) the realisticness of the reconstructed datasets.
    \vspace{-0.cm}
\end{enumerate}
\noindent\textbf{Model training:} We perform a user-based split between training and test datasets per province. Since the use cases involve shuffling and matching traffic and mobility sequences across users, the test dataset is restricted to users with complete sequences (i.e., no missing data). Users with incomplete sequences are included in the training dataset. Table \ref{tab:data-description} provides the number of users in the test dataset for each province, used for the inference cases. The training process focuses on constructing a Markov model by capturing the states and transitions exhibited by users in the training dataset. 

\vspace{0.13cm}
\noindent\textbf{Performance results:}\\
For the first use case, Fig. \ref{fig:mobtotra_usecase1} shows the accuracy of predicting users' mobility behavior from traffic patterns across different provinces. The overall mobility inference accuracy ranges from 94.3\% in Santiago, to 96.8\% in Bio-Bio, with minimal deviation across users. At the feature level, the model performs best in predicting repetitiveness, followed by stationarity and distance.  In contrast, traffic category inference from mobility behavior shows lower performance, with accuracies of of $41.7\% \pm 34.4$, $43.9\% \pm 31.2$, $52.4\% \pm 39.0$, and $36.8\% \pm 29.9$ in Santiago, Elqui, Bio-Bio, and Copiapo, respectively. 
\greybox{\textbf{Takeaway 4:} \textit{The richer dependencies in mobility patterns make it easier to infer user's mobility from traffic behavior than the contrary.}}

The second use case's results show a significantly higher likelihood for the regular dataset compared to the shuffled one, measured using the Hellinger distance. Fig. \ref{fig:hdis-intra} illustrates the best scores, ranging from 0.46 in Bio-Bio to 0.62 in Santiago, particularly when the empirical parameter $\alpha = 0$. The \textit{default} version incorporates both hour and minute time features, while the \textit{time-hour} version uses only the hour of the day. As $\alpha$ increases, the model’s ability to differentiate between behaviors diminishes, indicating that the rate of valid transitions in a sequence is key to effective modeling. The \textit{time-hour} version consistently performs better, suggesting improved generalization when time is more flexibly defined. The high Hellinger distances shown in Fig. \ref{fig:hdis-inter} across provinces validate the model’s capability to generalize across different regions, with some provinces like Bio-Bio exhibiting even better inference performance. Overall, the model achieves a near-1 likelihood distribution for regular behaviors, as depicted in Fig. \ref{fig:inter_distributions}. 
\greybox{\textbf{Takeaway 5:} \textit{The rate of valid transitions within a user behavior sequence serves as a critical and generalizable metric for differentiating between realistic and unrealistic mobility and traffic patterns.}}

In the third use case, we evaluate the model’s ability to accurately match mobility and traffic data across provinces using the \textit{time-hour} model with $\alpha=0$. Table \ref{tab:inference3} shows the matching accuracy, where Santiago's subset of 1000 users is sampled in different ways: (i) \textit{distrib}, preserving the original dataset’s mobility profile distribution, (ii) \textit{routiner}, featuring only routine-profile users, (iii) \textit{regular}, for regular-profile users, (iv) \textit{scouter}, and (v) \textit{uniform}, balancing all three profiles. Results show high Top-1 accuracy, ranging from 46\% in Bio-Bio to 66\% in Santiago, indicating significant performance for correctly matching users from among at least 1000 possible candidates. The table further highlights higher accuracy in the first 5\% to 20\% range. We also assess the realisticness of the model’s matching by comparing it with the ground truth. Fig. \ref{fig:max_likelihood} demonstrates that the ground-truth matches nearly always have the highest likelihood or, in cases where they don’t, a likelihood very close to the highest. This validates that the selected match is statistically realistic. Additionally, Fig. \ref{fig:max_hamming_dis} shows a negligible Hamming distance between the model's selected match and the ground truth, further validating the realistic nature of the model’s matching process.
\greybox{\textbf{Takeaway 6:} \textit{Modeling traffic-mobility interactions enables the creation of realistic, integrated datasets, combining user-level patterns from multiple sources, and facilitating flexible synthetic datasets.}}


\section{Conclusion}
\label{sec:conclusion}
This study introduced a novel approach to understanding the intricate relationship between individual traffic and mobility behaviors using eXtended Data Records (XDRs). By analyzing 13 user-level and time-level features and through a novel behavior-based user modeling, we built a Markov model capable of inferring mobility from traffic behavior and vice versa. Our findings show that mobility patterns are easier to predict from traffic data, while the rate of valid transitions serves as a key indicator for distinguishing realistic behavior sequences. Additionally, the model's ability to generalize across diverse urban contexts validates its robustness. This work paves the way for creating realistic, flexible synthetic datasets by integrating multi-sourced user behaviors.

\section{Acknowledgments}
This work was partially supported by CAPES-STIC-AMSUD 22-STIC-07 LINT project.
The authors also acknowledge financial support the Mob Sci-Dat Factory (ANR-23-PEMO-0004) project. 

\bibliographystyle{ACM-Reference-Format}
\bibliography{references}


\begin{thebibliography}{28}


\ifx \showCODEN    \undefined \def \showCODEN     #1{\unskip}     \fi
\ifx \showDOI      \undefined \def \showDOI       #1{#1}\fi
\ifx \showISBNx    \undefined \def \showISBNx     #1{\unskip}     \fi
\ifx \showISBNxiii \undefined \def \showISBNxiii  #1{\unskip}     \fi
\ifx \showISSN     \undefined \def \showISSN      #1{\unskip}     \fi
\ifx \showLCCN     \undefined \def \showLCCN      #1{\unskip}     \fi
\ifx \shownote     \undefined \def \shownote      #1{#1}          \fi
\ifx \showarticletitle \undefined \def \showarticletitle #1{#1}   \fi
\ifx \showURL      \undefined \def \showURL       {\relax}        \fi
\providecommand\bibfield[2]{#2}
\providecommand\bibinfo[2]{#2}
\providecommand\natexlab[1]{#1}
\providecommand\showeprint[2][]{arXiv:#2}

\bibitem[Aceto et~al\mbox{.}(2021)]%
        {Aceto:2021}
\bibfield{author}{\bibinfo{person}{Giuseppe Aceto}, \bibinfo{person}{Giampaolo Bovenzi}, \bibinfo{person}{Domenico Ciuonzo}, \bibinfo{person}{Antonio Montieri}, \bibinfo{person}{Valerio Persico}, {and} \bibinfo{person}{Antonio Pescapé}.} \bibinfo{year}{2021}\natexlab{}.
\newblock \showarticletitle{Characterization and Prediction of Mobile-App Traffic Using Markov Modeling}.
\newblock \bibinfo{journal}{\emph{IEEE Transactions on Network and Service Management}} \bibinfo{volume}{18}, \bibinfo{number}{1} (\bibinfo{year}{2021}), \bibinfo{pages}{907--925}.
\newblock
\urldef\tempurl%
\url{https://doi.org/10.1109/TNSM.2021.3051381}
\showDOI{\tempurl}


\bibitem[Alipour et~al\mbox{.}(2018)]%
        {Alipour:2018}
\bibfield{author}{\bibinfo{person}{Babak Alipour}, \bibinfo{person}{Leonardo Tonetto}, \bibinfo{person}{Aaron~Yi Ding}, \bibinfo{person}{Roozbeh Ketabi}, \bibinfo{person}{Jörg Ott}, {and} \bibinfo{person}{Ahmed Helmy}.} \bibinfo{year}{2018}\natexlab{}.
\newblock \showarticletitle{Flutes vs. Cellos: Analyzing Mobility-Traffic Correlations in Large WLAN Traces}. In \bibinfo{booktitle}{\emph{IEEE INFOCOM 2018 - IEEE Conference on Computer Communications}}. \bibinfo{pages}{1637--1645}.
\newblock
\urldef\tempurl%
\url{https://doi.org/10.1109/INFOCOM.2018.8486360}
\showDOI{\tempurl}


\bibitem[Amichi et~al\mbox{.}(2020)]%
        {Amichi:2020}
\bibfield{author}{\bibinfo{person}{Licia Amichi}, \bibinfo{person}{Aline~Carneiro Viana}, \bibinfo{person}{Mark Crovella}, {and} \bibinfo{person}{Antonio~A.F. Loureiro}.} \bibinfo{year}{2020}\natexlab{}.
\newblock \showarticletitle{Understanding individuals' proclivity for novelty seeking}. In \bibinfo{booktitle}{\emph{Proceedings of the 28th International Conference on Advances in Geographic Information Systems}} (Seattle, WA, USA) \emph{(\bibinfo{series}{SIGSPATIAL '20})}. \bibinfo{publisher}{Association for Computing Machinery}, \bibinfo{address}{New York, NY, USA}, \bibinfo{pages}{314–324}.
\newblock
\showISBNx{9781450380195}
\urldef\tempurl%
\url{https://doi.org/10.1145/3397536.3422248}
\showDOI{\tempurl}


\bibitem[Chen et~al\mbox{.}(2015)]%
        {Chen:2015}
\bibfield{author}{\bibinfo{person}{Xiaming Chen}, \bibinfo{person}{Yaohui Jin}, \bibinfo{person}{Siwei Qiang}, \bibinfo{person}{Weisheng Hu}, {and} \bibinfo{person}{Kaida Jiang}.} \bibinfo{year}{2015}\natexlab{}.
\newblock \showarticletitle{Analyzing and modeling spatio-temporal dependence of cellular traffic at city scale}. In \bibinfo{booktitle}{\emph{2015 IEEE International Conference on Communications (ICC)}}. \bibinfo{pages}{3585--3591}.
\newblock
\urldef\tempurl%
\url{https://doi.org/10.1109/ICC.2015.7248881}
\showDOI{\tempurl}


\bibitem[{De Domenico} et~al\mbox{.}(2013)]%
        {DeDomenico:2013}
\bibfield{author}{\bibinfo{person}{Manlio {De Domenico}}, \bibinfo{person}{Antonio Lima}, {and} \bibinfo{person}{Mirco Musolesi}.} \bibinfo{year}{2013}\natexlab{}.
\newblock \showarticletitle{Interdependence and predictability of human mobility and social interactions}.
\newblock \bibinfo{journal}{\emph{Pervasive and Mobile Computing}} \bibinfo{volume}{9}, \bibinfo{number}{6} (\bibinfo{year}{2013}), \bibinfo{pages}{798--807}.
\newblock
\showISSN{1574-1192}
\urldef\tempurl%
\url{https://doi.org/10.1016/j.pmcj.2013.07.008}
\showDOI{\tempurl}
\newblock
\shownote{Mobile Data Challenge}.


\bibitem[Demissie et~al\mbox{.}(2013)]%
        {Merkebe:2013}
\bibfield{author}{\bibinfo{person}{Merkebe~Getachew Demissie}, \bibinfo{person}{Gonçalo~Homem de Almeida~Correia}, {and} \bibinfo{person}{Carlos Bento}.} \bibinfo{year}{2013}\natexlab{}.
\newblock \showarticletitle{Exploring cellular network handover information for urban mobility analysis}.
\newblock \bibinfo{journal}{\emph{Journal of Transport Geography}}  \bibinfo{volume}{31} (\bibinfo{year}{2013}), \bibinfo{pages}{164--170}.
\newblock
\showISSN{0966-6923}
\urldef\tempurl%
\url{https://doi.org/10.1016/j.jtrangeo.2013.06.016}
\showDOI{\tempurl}


\bibitem[González et~al\mbox{.}(2008)]%
        {Gonzalez:2008}
\bibfield{author}{\bibinfo{person}{Marta~C. González}, \bibinfo{person}{César~A. Hidalgo}, {and} \bibinfo{person}{Albert-László Barabási}.} \bibinfo{year}{2008}\natexlab{}.
\newblock \showarticletitle{Understanding individual human mobility patterns}.
\newblock \bibinfo{journal}{\emph{Nature}} \bibinfo{volume}{453}, \bibinfo{number}{7196} (\bibinfo{year}{2008}), \bibinfo{pages}{779--782}.
\newblock
\urldef\tempurl%
\url{https://doi.org/10.1038/nature06958}
\showDOI{\tempurl}


\bibitem[Jiang et~al\mbox{.}(2013)]%
        {Jiang:2013}
\bibfield{author}{\bibinfo{person}{Shan Jiang}, \bibinfo{person}{Gaston~A Fiore}, \bibinfo{person}{Yingxiang Yang}, \bibinfo{person}{Joseph Ferreira~Jr}, \bibinfo{person}{Emilio Frazzoli}, {and} \bibinfo{person}{Marta~C Gonz{\'a}lez}.} \bibinfo{year}{2013}\natexlab{}.
\newblock \showarticletitle{A review of urban computing for mobile phone traces: {Current} methods, challenges and opportunities}. In \bibinfo{booktitle}{\emph{Proceedings of the 2nd ACM SIGKDD international workshop on Urban Computing}}. ACM, \bibinfo{pages}{2}.
\newblock


\bibitem[{Keramat Jahromi} et~al\mbox{.}(2016)]%
        {Keramat:2016}
\bibfield{author}{\bibinfo{person}{Karim {Keramat Jahromi}}, \bibinfo{person}{Matteo Zignani}, \bibinfo{person}{Sabrina Gaito}, {and} \bibinfo{person}{Gian~Paolo Rossi}.} \bibinfo{year}{2016}\natexlab{}.
\newblock \showarticletitle{Simulating human mobility patterns in urban areas}.
\newblock \bibinfo{journal}{\emph{Simulation Modelling Practice and Theory}}  \bibinfo{volume}{62} (\bibinfo{year}{2016}), \bibinfo{pages}{137--156}.
\newblock
\showISSN{1569-190X}
\urldef\tempurl%
\url{https://doi.org/10.1016/j.simpat.2015.12.002}
\showDOI{\tempurl}


\bibitem[Kobayashi et~al\mbox{.}(2023)]%
        {Kobayashi:2023}
\bibfield{author}{\bibinfo{person}{Akihiro Kobayashi}, \bibinfo{person}{Naoto Takeda}, \bibinfo{person}{Yudai Yamazaki}, {and} \bibinfo{person}{Daisuke Kamisaka}.} \bibinfo{year}{2023}\natexlab{}.
\newblock \showarticletitle{Modeling and generating human mobility trajectories using transformer with day encoding}. In \bibinfo{booktitle}{\emph{Proceedings of the 1st International Workshop on the Human Mobility Prediction Challenge}} (Hamburg, Germany) \emph{(\bibinfo{series}{HuMob-Challenge '23})}. \bibinfo{publisher}{Association for Computing Machinery}, \bibinfo{address}{New York, NY, USA}, \bibinfo{pages}{7–10}.
\newblock
\showISBNx{9798400703560}
\urldef\tempurl%
\url{https://doi.org/10.1145/3615894.3628506}
\showDOI{\tempurl}


\bibitem[Kontoyiannis et~al\mbox{.}(1998)]%
        {Kontoyiannis:1998}
\bibfield{author}{\bibinfo{person}{I. Kontoyiannis}, \bibinfo{person}{P.H. Algoet}, \bibinfo{person}{Yu.M. Suhov}, {and} \bibinfo{person}{A.J. Wyner}.} \bibinfo{year}{1998}\natexlab{}.
\newblock \showarticletitle{Nonparametric entropy estimation for stationary processes and random fields, with applications to English text}.
\newblock \bibinfo{journal}{\emph{IEEE Transactions on Information Theory}} \bibinfo{volume}{44}, \bibinfo{number}{3} (\bibinfo{year}{1998}), \bibinfo{pages}{1319--1327}.
\newblock
\urldef\tempurl%
\url{https://doi.org/10.1109/18.669425}
\showDOI{\tempurl}


\bibitem[Koszowski et~al\mbox{.}(2019)]%
        {Koszowski:2019}
\bibfield{author}{\bibinfo{person}{Caroline Koszowski}, \bibinfo{person}{Regine Gerike}, \bibinfo{person}{Stefan Hubrich}, \bibinfo{person}{Thomas G{\"o}tschi}, \bibinfo{person}{Maria Pohle}, {and} \bibinfo{person}{Rico Wittwer}.} \bibinfo{year}{2019}\natexlab{}.
\newblock \bibinfo{booktitle}{\emph{Active Mobility: Bringing Together Transport Planning, Urban Planning, and Public Health}}.
\newblock \bibinfo{publisher}{Springer International Publishing}, \bibinfo{address}{Cham}, \bibinfo{pages}{149--171}.
\newblock
\showISBNx{978-3-319-99756-8}
\urldef\tempurl%
\url{https://doi.org/10.1007/978-3-319-99756-8_11}
\showDOI{\tempurl}


\bibitem[Kouam et~al\mbox{.}(2023)]%
        {Kouam:2023}
\bibfield{author}{\bibinfo{person}{Anne~Josiane Kouam}, \bibinfo{person}{Aline Carneiro~Viana}, {and} \bibinfo{person}{Alain Tchana}.} \bibinfo{year}{2023}\natexlab{}.
\newblock \showarticletitle{LSTM-based generation of cellular network traffic}. In \bibinfo{booktitle}{\emph{2023 IEEE Wireless Communications and Networking Conference (WCNC)}}. \bibinfo{pages}{1--6}.
\newblock
\urldef\tempurl%
\url{https://doi.org/10.1109/WCNC55385.2023.10119094}
\showDOI{\tempurl}


\bibitem[Mucceli et~al\mbox{.}(2016)]%
        {mucceli:2016}
\bibfield{author}{\bibinfo{person}{Eduardo Mucceli}, \bibinfo{person}{Aline Carneiro~Viana}, \bibinfo{person}{Carlos Sarraute}, \bibinfo{person}{Jorge Brea}, {and} \bibinfo{person}{Jos{\'e}~Ignacio Alvarez-Hamelin}.} \bibinfo{year}{2016}\natexlab{}.
\newblock \showarticletitle{{On the Regularity of Human Mobility}}.
\newblock \bibinfo{journal}{\emph{{Pervasive and Mobile Computing}}} (\bibinfo{date}{Dec.} \bibinfo{year}{2016}).
\newblock
\urldef\tempurl%
\url{https://inria.hal.science/hal-01367825}
\showURL{%
\tempurl}


\bibitem[Mucelli Rezende~Oliveira et~al\mbox{.}(2015)]%
        {Eduardo:2015}
\bibfield{author}{\bibinfo{person}{Eduardo Mucelli Rezende~Oliveira}, \bibinfo{person}{Aline Carneiro~Viana}, \bibinfo{person}{Kolar~Purushothama Naveen}, {and} \bibinfo{person}{Carlos Sarraute}.} \bibinfo{year}{2015}\natexlab{}.
\newblock \showarticletitle{{Measurement-driven mobile data traffic modeling in a large metropolitan area}}. In \bibinfo{booktitle}{\emph{{PerCom 2015- 13th Conference on Pervasive Computing and Communications}}}. \bibinfo{publisher}{{IEEE}}, \bibinfo{address}{St. Louis, Missouri, United States}.
\newblock
\urldef\tempurl%
\url{https://inria.hal.science/hal-01089434}
\showURL{%
\tempurl}


\bibitem[Naboulsi et~al\mbox{.}(2014)]%
        {Naboulsi:2014}
\bibfield{author}{\bibinfo{person}{Diala Naboulsi}, \bibinfo{person}{Razvan Stanica}, {and} \bibinfo{person}{Marco Fiore}.} \bibinfo{year}{2014}\natexlab{}.
\newblock \showarticletitle{Classifying call profiles in large-scale mobile traffic datasets}. In \bibinfo{booktitle}{\emph{IEEE INFOCOM 2014 - IEEE Conference on Computer Communications}}. \bibinfo{pages}{1806--1814}.
\newblock
\urldef\tempurl%
\url{https://doi.org/10.1109/INFOCOM.2014.6848119}
\showDOI{\tempurl}


\bibitem[Ozturk et~al\mbox{.}(2021)]%
        {Ozturk:2021}
\bibfield{author}{\bibinfo{person}{Metin Ozturk}, \bibinfo{person}{Attai~Ibrahim Abubakar}, \bibinfo{person}{João Pedro~Battistella Nadas}, \bibinfo{person}{Rao Naveed~Bin Rais}, \bibinfo{person}{Sajjad Hussain}, {and} \bibinfo{person}{Muhammad~Ali Imran}.} \bibinfo{year}{2021}\natexlab{}.
\newblock \showarticletitle{Energy Optimization in Ultra-Dense Radio Access Networks via Traffic-Aware Cell Switching}.
\newblock \bibinfo{journal}{\emph{IEEE Transactions on Green Communications and Networking}} \bibinfo{volume}{5}, \bibinfo{number}{2} (\bibinfo{year}{2021}), \bibinfo{pages}{832--845}.
\newblock
\urldef\tempurl%
\url{https://doi.org/10.1109/TGCN.2021.3056235}
\showDOI{\tempurl}


\bibitem[Pappalardo et~al\mbox{.}(2015)]%
        {Pappalardo:2015}
\bibfield{author}{\bibinfo{person}{Luca Pappalardo}, \bibinfo{person}{Filippo Simini}, \bibinfo{person}{Salvatore Rinzivillo}, \bibinfo{person}{Dino Pedreschi}, \bibinfo{person}{Fosca Giannotti}, {and} \bibinfo{person}{Albert-László Barabási}.} \bibinfo{year}{2015}\natexlab{}.
\newblock \showarticletitle{Returners and explorers dichotomy in human mobility}.
\newblock \bibinfo{journal}{\emph{Nature Communications}}  \bibinfo{volume}{6} (\bibinfo{year}{2015}), \bibinfo{pages}{8166}.
\newblock
\showISSN{2041-1723}
\urldef\tempurl%
\url{https://doi.org/10.1038/ncomms9166}
\showDOI{\tempurl}


\bibitem[Paul et~al\mbox{.}(2011)]%
        {Paul:2011}
\bibfield{author}{\bibinfo{person}{Utpal Paul}, \bibinfo{person}{Anand~Prabhu Subramanian}, \bibinfo{person}{Milind~Madhav Buddhikot}, {and} \bibinfo{person}{Samir~R. Das}.} \bibinfo{year}{2011}\natexlab{}.
\newblock \showarticletitle{Understanding traffic dynamics in cellular data networks}. In \bibinfo{booktitle}{\emph{2011 Proceedings IEEE INFOCOM}}. \bibinfo{pages}{882--890}.
\newblock
\urldef\tempurl%
\url{https://doi.org/10.1109/INFCOM.2011.5935313}
\showDOI{\tempurl}


\bibitem[Rhee et~al\mbox{.}(2011)]%
        {Rhee:2011}
\bibfield{author}{\bibinfo{person}{Injong Rhee}, \bibinfo{person}{Minsu Shin}, \bibinfo{person}{Seongik Hong}, \bibinfo{person}{Kyunghan Lee}, \bibinfo{person}{Seong~Joon Kim}, {and} \bibinfo{person}{Song Chong}.} \bibinfo{year}{2011}\natexlab{}.
\newblock \showarticletitle{On the Levy-Walk Nature of Human Mobility}.
\newblock \bibinfo{journal}{\emph{IEEE/ACM Transactions on Networking}} \bibinfo{volume}{19}, \bibinfo{number}{3} (\bibinfo{year}{2011}), \bibinfo{pages}{630--643}.
\newblock
\urldef\tempurl%
\url{https://doi.org/10.1109/TNET.2011.2120618}
\showDOI{\tempurl}


\bibitem[Scherrer et~al\mbox{.}(2018)]%
        {Scherrer:2018}
\bibfield{author}{\bibinfo{person}{Luca Scherrer}, \bibinfo{person}{Martin Tomko}, \bibinfo{person}{Peter Ranacher}, {and} \bibinfo{person}{Robert Weibel}.} \bibinfo{year}{2018}\natexlab{}.
\newblock \showarticletitle{Travelers or locals? Identifying meaningful sub-populations from human movement data in the absence of ground truth}.
\newblock \bibinfo{journal}{\emph{EPJ Data Science}} \bibinfo{volume}{7}, \bibinfo{number}{1} (\bibinfo{year}{2018}), \bibinfo{pages}{19}.
\newblock
\showISSN{2193-1127}
\urldef\tempurl%
\url{https://doi.org/10.1140/epjds/s13688-018-0147-7}
\showDOI{\tempurl}


\bibitem[Trinh et~al\mbox{.}(2017)]%
        {Trinh:2017}
\bibfield{author}{\bibinfo{person}{Hoang~Duy Trinh}, \bibinfo{person}{Nicola Bui}, \bibinfo{person}{Joerg Widmer}, \bibinfo{person}{Lorenza Giupponi}, {and} \bibinfo{person}{Paolo Dini}.} \bibinfo{year}{2017}\natexlab{}.
\newblock \showarticletitle{Analysis and modeling of mobile traffic using real traces}. In \bibinfo{booktitle}{\emph{2017 IEEE 28th Annual International Symposium on Personal, Indoor, and Mobile Radio Communications (PIMRC)}}. \bibinfo{pages}{1--6}.
\newblock
\urldef\tempurl%
\url{https://doi.org/10.1109/PIMRC.2017.8292200}
\showDOI{\tempurl}


\bibitem[Wang et~al\mbox{.}(2019a)]%
        {Wang_Jinzhong:2019}
\bibfield{author}{\bibinfo{person}{Jinzhong Wang}, \bibinfo{person}{Xiangjie Kong}, \bibinfo{person}{Feng Xia}, {and} \bibinfo{person}{Lijun Sun}.} \bibinfo{year}{2019}\natexlab{a}.
\newblock \showarticletitle{Urban Human Mobility: Data-Driven Modeling and Prediction}.
\newblock \bibinfo{journal}{\emph{SIGKDD Explor. Newsl.}} \bibinfo{volume}{21}, \bibinfo{number}{1} (\bibinfo{date}{may} \bibinfo{year}{2019}), \bibinfo{pages}{1–19}.
\newblock
\showISSN{1931-0145}
\urldef\tempurl%
\url{https://doi.org/10.1145/3331651.3331653}
\showDOI{\tempurl}


\bibitem[Wang et~al\mbox{.}(2019b)]%
        {Wang:2019}
\bibfield{author}{\bibinfo{person}{Xu Wang}, \bibinfo{person}{Zimu Zhou}, \bibinfo{person}{Fu Xiao}, \bibinfo{person}{Kai Xing}, \bibinfo{person}{Zheng Yang}, \bibinfo{person}{Yunhao Liu}, {and} \bibinfo{person}{Chunyi Peng}.} \bibinfo{year}{2019}\natexlab{b}.
\newblock \showarticletitle{Spatio-Temporal Analysis and Prediction of Cellular Traffic in Metropolis}.
\newblock \bibinfo{journal}{\emph{IEEE Transactions on Mobile Computing}} \bibinfo{volume}{18}, \bibinfo{number}{9} (\bibinfo{year}{2019}), \bibinfo{pages}{2190--2202}.
\newblock
\urldef\tempurl%
\url{https://doi.org/10.1109/TMC.2018.2870135}
\showDOI{\tempurl}


\bibitem[Wu et~al\mbox{.}(2018a)]%
        {Wu_Jing:2018}
\bibfield{author}{\bibinfo{person}{Jing Wu}, \bibinfo{person}{Ming Zeng}, \bibinfo{person}{Xinlei Chen}, \bibinfo{person}{Yong Li}, {and} \bibinfo{person}{Depeng Jin}.} \bibinfo{year}{2018}\natexlab{a}.
\newblock \showarticletitle{Characterizing and Predicting Individual Traffic Usage of Mobile Application in Cellular Network}. In \bibinfo{booktitle}{\emph{Proceedings of the 2018 ACM International Joint Conference and 2018 International Symposium on Pervasive and Ubiquitous Computing and Wearable Computers}} (Singapore, Singapore) \emph{(\bibinfo{series}{UbiComp '18})}. \bibinfo{publisher}{Association for Computing Machinery}, \bibinfo{address}{New York, NY, USA}, \bibinfo{pages}{852–861}.
\newblock
\showISBNx{9781450359665}
\urldef\tempurl%
\url{https://doi.org/10.1145/3267305.3274173}
\showDOI{\tempurl}


\bibitem[Wu et~al\mbox{.}(2018b)]%
        {Jing:2018}
\bibfield{author}{\bibinfo{person}{Jing Wu}, \bibinfo{person}{Ming Zeng}, \bibinfo{person}{Xinlei Chen}, \bibinfo{person}{Yong Li}, {and} \bibinfo{person}{Depeng Jin}.} \bibinfo{year}{2018}\natexlab{b}.
\newblock \showarticletitle{Characterizing and Predicting Individual Traffic Usage of Mobile Application in Cellular Network}. In \bibinfo{booktitle}{\emph{Proceedings of the 2018 ACM International Joint Conference and 2018 International Symposium on Pervasive and Ubiquitous Computing and Wearable Computers}} (Singapore, Singapore) \emph{(\bibinfo{series}{UbiComp '18})}. \bibinfo{publisher}{Association for Computing Machinery}, \bibinfo{address}{New York, NY, USA}, \bibinfo{pages}{852–861}.
\newblock
\showISBNx{9781450359665}
\urldef\tempurl%
\url{https://doi.org/10.1145/3267305.3274173}
\showDOI{\tempurl}


\bibitem[Xu et~al\mbox{.}(2017)]%
        {Xu:2017}
\bibfield{author}{\bibinfo{person}{Fengli Xu}, \bibinfo{person}{Yong Li}, \bibinfo{person}{Huandong Wang}, \bibinfo{person}{Pengyu Zhang}, {and} \bibinfo{person}{Depeng Jin}.} \bibinfo{year}{2017}\natexlab{}.
\newblock \showarticletitle{Understanding Mobile Traffic Patterns of Large Scale Cellular Towers in Urban Environment}.
\newblock \bibinfo{journal}{\emph{IEEE/ACM Transactions on Networking}} \bibinfo{volume}{25}, \bibinfo{number}{2} (\bibinfo{year}{2017}), \bibinfo{pages}{1147--1161}.
\newblock


\bibitem[Xu et~al\mbox{.}(2021)]%
        {Xu_Kai:2021}
\bibfield{author}{\bibinfo{person}{Kai Xu}, \bibinfo{person}{Rajkarn Singh}, \bibinfo{person}{Marco Fiore}, \bibinfo{person}{Mahesh~K. Marina}, \bibinfo{person}{Hakan Bilen}, \bibinfo{person}{Muhammad Usama}, \bibinfo{person}{Howard Benn}, {and} \bibinfo{person}{Cezary Ziemlicki}.} \bibinfo{year}{2021}\natexlab{}.
\newblock \showarticletitle{SpectraGAN: spectrum based generation of city scale spatiotemporal mobile network traffic data}. In \bibinfo{booktitle}{\emph{Proceedings of the 17th International Conference on Emerging Networking EXperiments and Technologies}} (Virtual Event, Germany) \emph{(\bibinfo{series}{CoNEXT '21})}. \bibinfo{publisher}{Association for Computing Machinery}, \bibinfo{address}{New York, NY, USA}, \bibinfo{pages}{243–258}.
\newblock
\showISBNx{9781450390989}
\urldef\tempurl%
\url{https://doi.org/10.1145/3485983.3494844}
\showDOI{\tempurl}


\end{thebibliography}

    
    



\appendix


\begin{figure}[h!]
    \centering
    \includegraphics[width=0.5\linewidth]{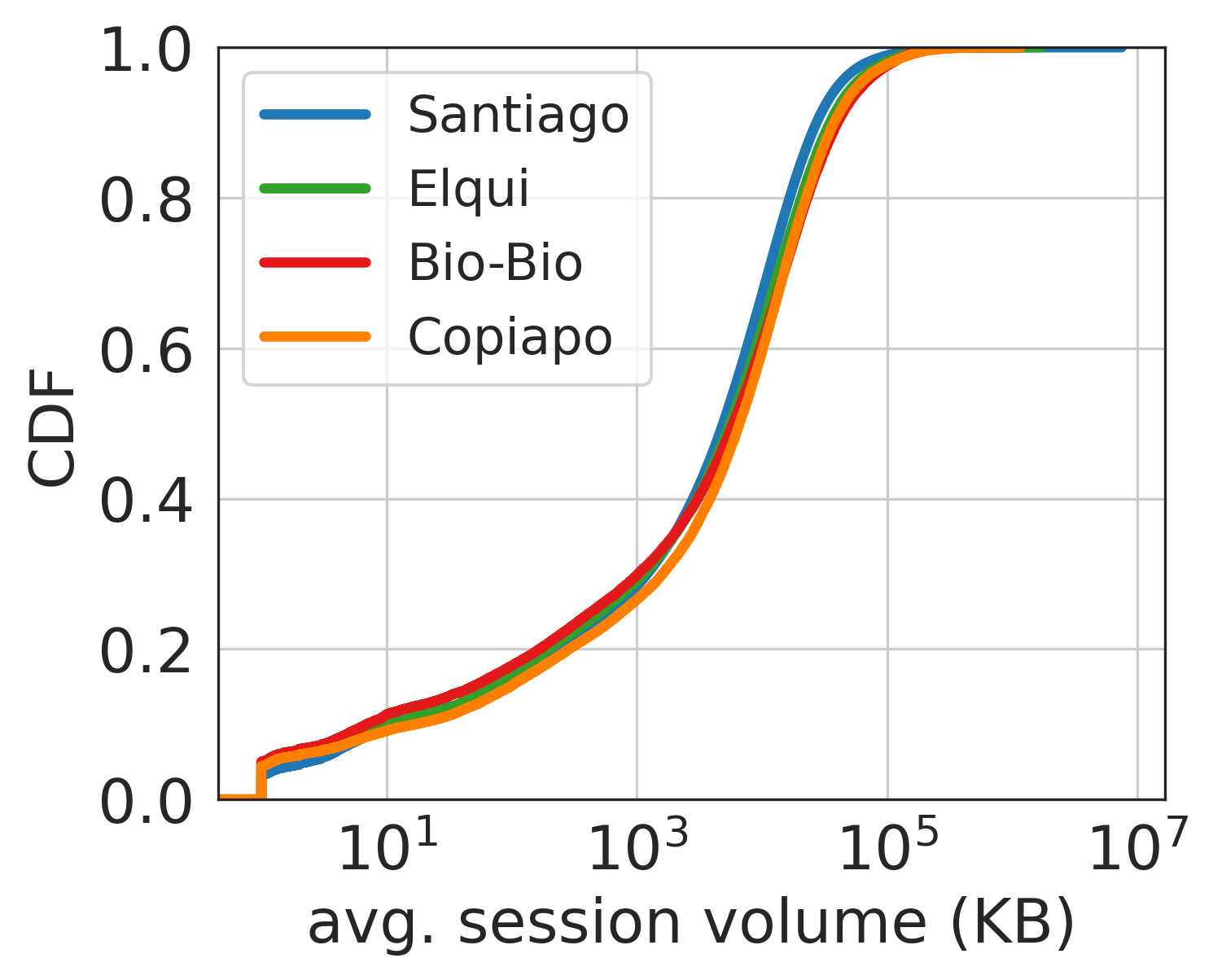}
    \Description{Mobile users' avg. traffic session volume distribution}
    \caption{Mobile users' avg. traffic session volume distribution}
    \label{fig:avg_volume}
\end{figure}

\begin{figure}[h!]
    \centering
    \includegraphics[width=\linewidth]{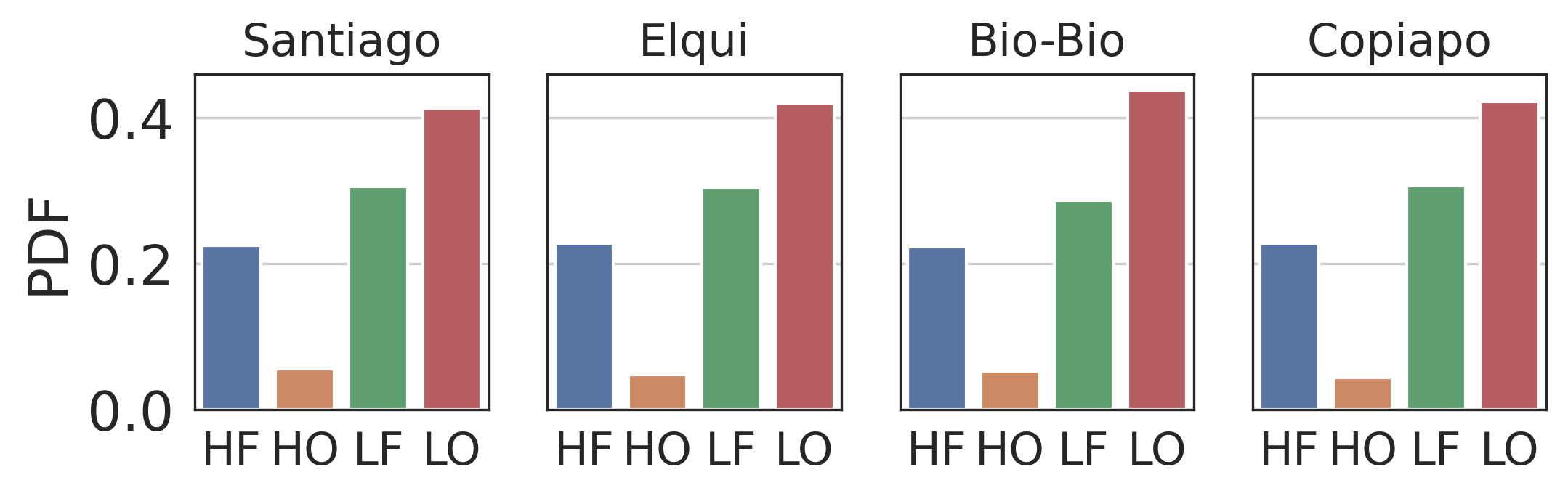}
    \Description{Distribution of users per traffic profile}
    \caption{Distribution of users per traffic profile}
    \label{fig:tra_profile_norm}
\end{figure}

\begin{figure}[h!]
    \centering
    \includegraphics[width=\linewidth]{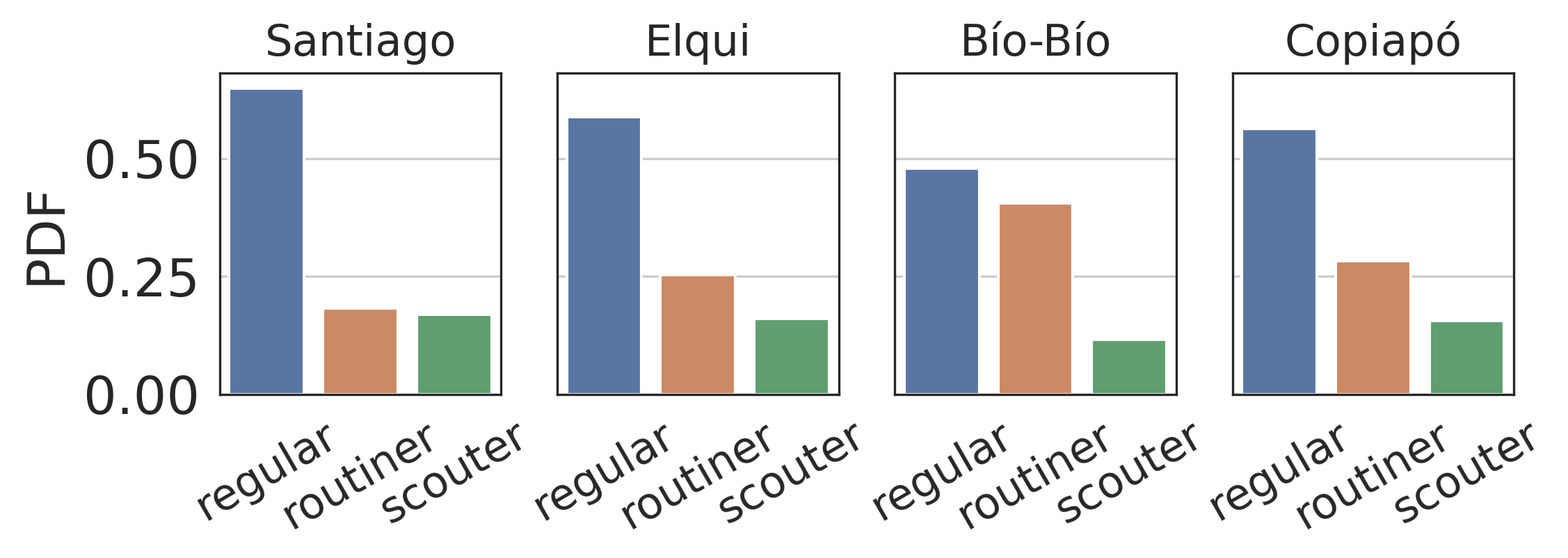}
    \Description{Distribution of users per mobility profile}
    \caption{Distribution of users per mobility profile}
    \label{fig:mob_profile_norm}
\end{figure}


\begin{figure*}[h!]
    \centering
    \includegraphics[width=0.7\linewidth]{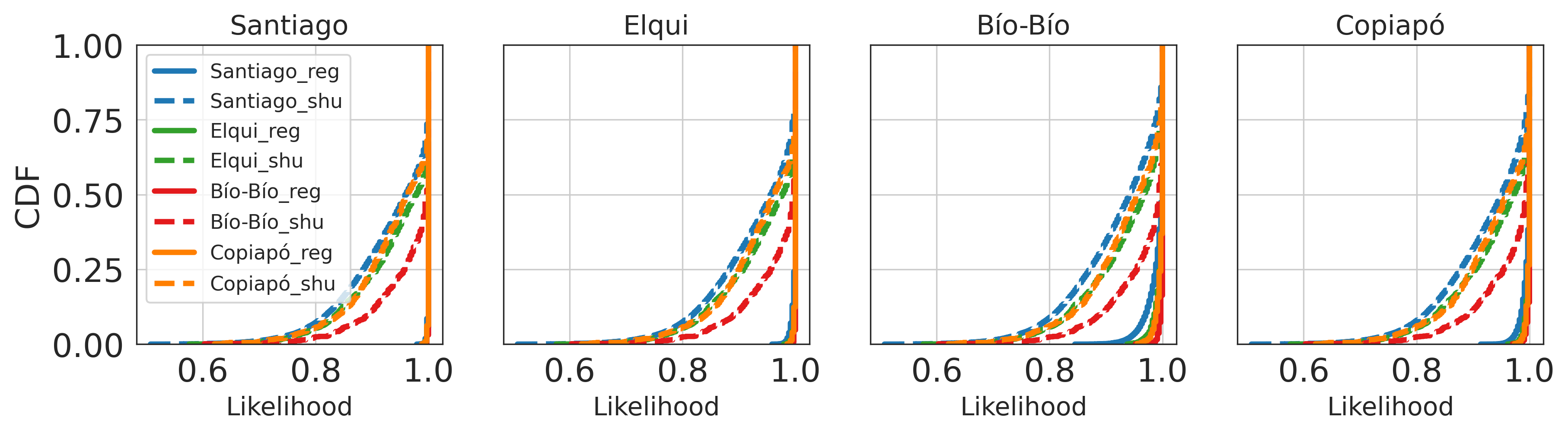}
    \Description{Regular and shuffled likelihood distributions with inter-province inference, \textit{time-hour} version, and $\alpha=0$}
    \caption{Regular and shuffled likelihood distributions with inter-province inference, \textit{time-hour} version, and $\alpha=0$}
    \label{fig:inter_distributions}
\end{figure*}

\begin{figure*}[h!]
    \centering
    \includegraphics[width=0.75\linewidth]{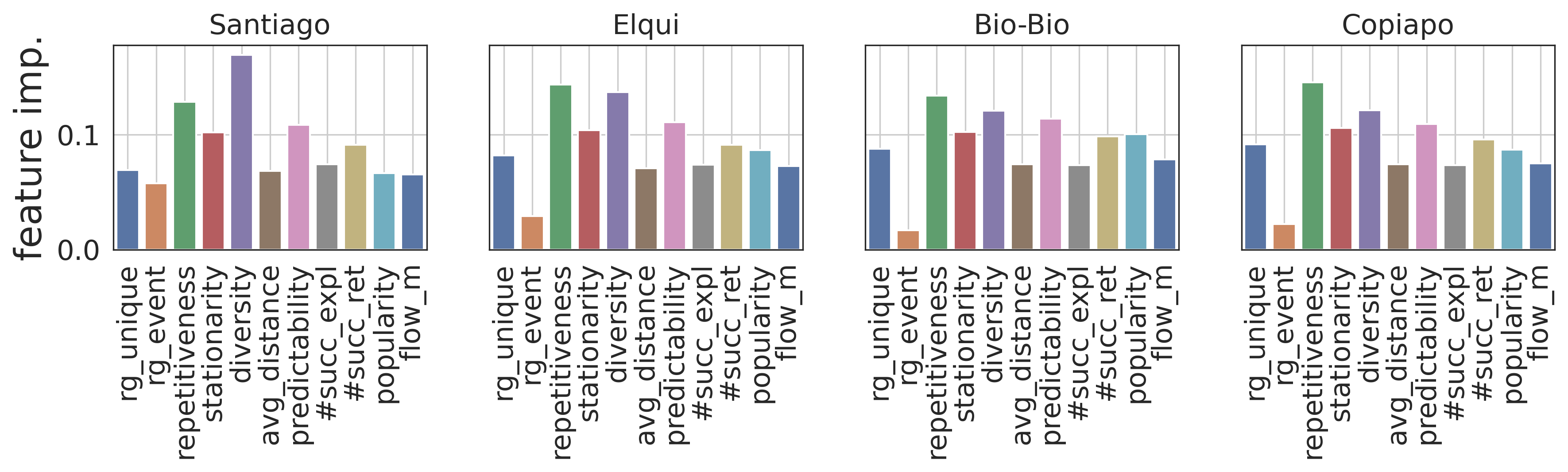}
    \Description{Importance of mobility features at explaining users' traffic profile}
    \caption{Importance of mobility features at explaining users' traffic profile}
    \label{fig:fi_mob_for_traffi}
\end{figure*}

\begin{figure*}[htbp]
    \centering
    \begin{subfigure}{0.6\linewidth}
        \centering
        \includegraphics[width=\linewidth]{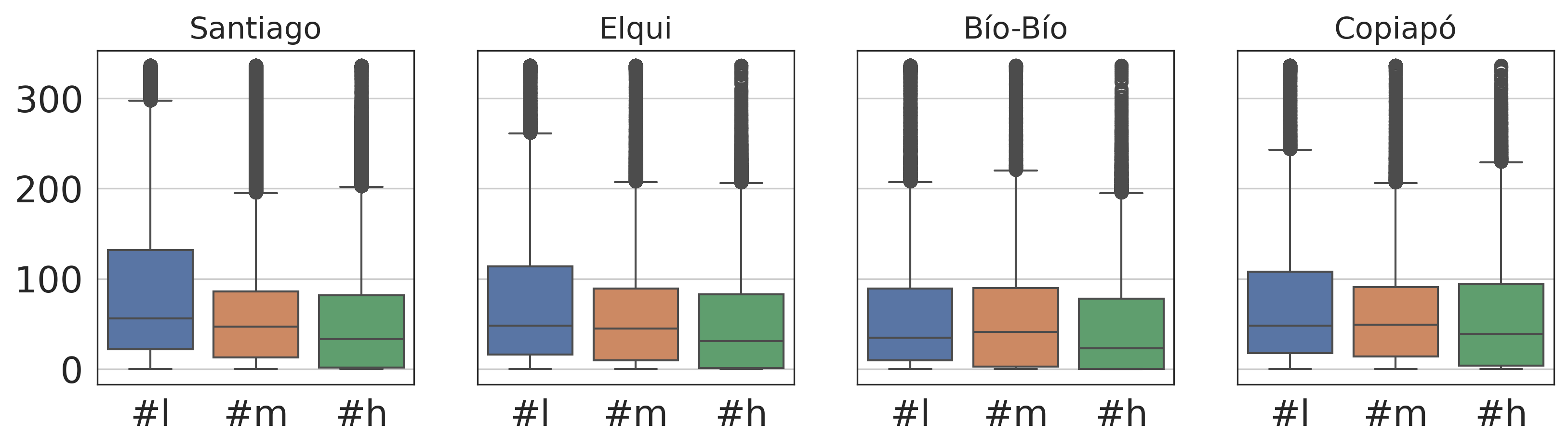}
        \Description{Distribution across of entire sequence}
        \caption{Distribution across of entire sequence}
        \label{fig:step_entire_distrib}
    \end{subfigure}
    \hfill
    \begin{subfigure}{0.3\linewidth}
        \centering
        \includegraphics[width=\linewidth]{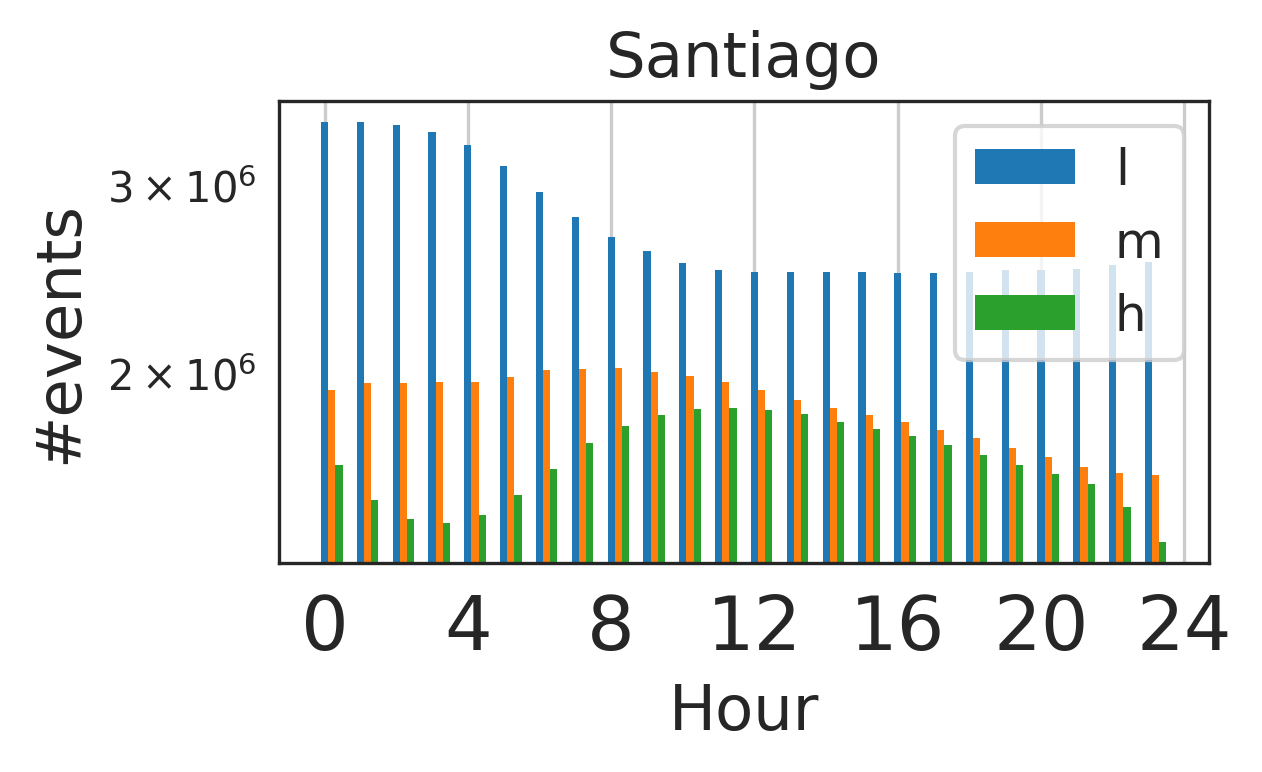}
        \Description{Hourly distribution in Santiago}
        \caption{Hourly distribution in Santiago}
        \label{fig:step_hourly_distrib}
    \end{subfigure}
    \caption{Distribution of users' number of events per traffic category}
    \label{fig:step_distrib}
\end{figure*}

\section{Feature importance extraction}
\label{app:feature_importance}
To identify the most significant mobility features influencing user traffic profiles (\textit{LO, LF, HO, HF}), we employ an Ensemble Trees Classifier. This approach leverages the strengths of multiple decision trees to enhance predictive accuracy and robustness. The methodology begins by inputting a set of mobility features derived from user behavior data, which serves as the basis for classifying traffic profiles.

The classifier processes this feature set through a training phase, where each tree in the ensemble learns to associate specific mobility patterns with the corresponding traffic profiles. After training, we assess feature importance using Gini importance, which measures the contribution of each feature to the model's predictive power. 

Gini importance is calculated by examining the reduction in Gini impurity, a metric that quantifies how often a feature correctly classifies instances. Specifically, when a feature is used to split data at a node in a decision tree, the Gini impurity decreases, indicating improved classification. The Gini importance of a feature is determined by averaging the impurity reduction across all trees in the ensemble. Higher Gini importance scores indicate that a feature has a more substantial impact on the model's ability to make accurate predictions.

The results of this assessment reveal a range of feature importances, scaled between 0 and 1, enabling us to rank mobility features based on their relevance to traffic profile classification as in Fig. \ref{fig:fi_mob_for_traffi}. Features with a relative higher importance scores indicate a stronger influence on the model's predictions, guiding further analysis and potential feature selection for enhancing the classification process.

\end{document}